%% file: main.tex
\documentclass[aps,pre,twocolumn,superscriptaddress,nofootinbib,longbibliography,floatfix]{revtex4-2}

\usepackage[T1]{fontenc}
\usepackage{amsmath,amssymb,amsthm}
\usepackage{bm}
\usepackage{graphicx}
\usepackage{booktabs}
\usepackage{tikz}
\usetikzlibrary{positioning,arrows.meta,calc}
\usepackage[colorlinks=true,linkcolor=blue!50!black,citecolor=blue!50!black,urlcolor=blue!50!black]{hyperref}

\theoremstyle{plain}
\newtheorem{theorem}{Theorem}
\newtheorem{proposition}{Proposition}
\newtheorem{lemma}{Lemma}
\newtheorem{corollary}{Corollary}
\theoremstyle{definition}
\newtheorem{definition}{Definition}
\newtheorem{example}{Example}
\theoremstyle{remark}
\newtheorem{remark}{Remark}

\newcommand{\Red}{\mathrm{Red}}
\newcommand{\Syn}{\mathrm{Syn}}
\newcommand{\Un}{\mathrm{Un}}
\newcommand{\pos}[1]{\left[#1\right]^{+}}
\newcommand{\dmax}{\delta_{\max}}
\newcommand{\Itwo}{C^{(2)}}
\newcommand{\Ithree}{I^{(3)}}
\newcommand{\Ptwo}{P^{(2)}}
\newcommand{\QCI}{Q_{\mathrm{CI}}}
\newcommand{\Fam}{\mathcal{Q}}
\newcommand{\Ftwo}{\mathcal{F}_{2}}
\newcommand{\DeltaP}{\Delta_{P}}
\newcommand{\QMI}{\mathcal{Q}_{\mathrm{MI}}}
\newcommand{\Idep}{I_{\mathrm{dep}}}
\newcommand{\Imin}{I_{\min}}
\newcommand{\Iccs}{I_{\mathrm{ccs}}}
\newcommand{\Isx}{I^{\mathrm{sx}}}
\newcommand{\Ipm}{I_{\pm}}
\newcommand{\Irr}{I_{\mathrm{rr}}}
\newcommand{\IGK}{I_{\wedge}}
\newcommand{\Ired}{I_{\mathrm{red}}}
\newcommand{\marg}{(\mathrm{SM})}%

\begin{document}

\title{Between Hi-Fi and MMI: the overlap of redundancy and synergy in partial information decomposition}

\author{Niels Van Santen}
\email[Corresponding author: ]{niels.vansanten@ugent.be}
\affiliation{Ghent University, Ghent, Belgium}
\author{Yves Rosseel}
\affiliation{Ghent University, Ghent, Belgium}
\author{Daniele Marinazzo}
\affiliation{Ghent University, Ghent, Belgium}

\date{\today}

\begin{abstract}
Partial information decomposition (PID) splits the information that two sources carry about a target into redundant, unique and synergistic parts, with nonnegative decompositions lying on a segment where redundancy measures are bounded by the positive part of the co-information and the smaller source--target information. We parameterize this segment by the overlap, the common excess of redundancy and synergy over their net balance, and show that for many of the named PID measures, their redundancy can be written as the largest positive part of the co-information over a set of reference distributions. This common form induces an ordering of those PIDs on the segment. With the smallest redundancy and synergy and the largest unique atoms of any nonnegative PID, the information-theoretic version of Hi-Fi, a feature-importance method, is the zero-overlap end of the segment for two sources, with the maximum overlap bound given by the minimum mutual information (MMI) PID. We introduce, by splitting the co-information into a pairwise maximum-entropy part and the third-order connected information, a PID that sits at zero overlap except when pairwise redundancy and triadic synergy coexist.
For three or more sources the bivariate results hold on every pair of merged source groups, and among all nonnegative solutions of the lattice equations MMI and Hi-Fi still give the smallest and largest unique atoms but zero overlap on all pairs at once is often unattainable. Two inequalities across pairs are necessary in general and sufficient for three sources. For identity targets zero overlap is possible exactly when the sources' information diagram is nonnegative. In random three- and four-source distributions zero overlap on all pairs fails in 21--100\% of the cases.
\end{abstract}

\maketitle

\input{sec1_intro}
\input{sec2_two}
\input{sec3_measures}
\input{sec4_triadic}
\input{sec6_mmi}
\input{sec7_multi}
\input{sec8_discussion}

\begin{acknowledgments}
This research was conducted with the support of the Special Research Fund (BOF/24J/2023/015).
\end{acknowledgments}

\section*{Data and code availability}
The code that reproduces all numerical results can be requested from the corresponding author and will be made publicly available upon publication.

\appendix
\input{appA}
\input{app_three}

\bibliography{references}

\end{document}

%% file: sec1_intro.tex
\section{Introduction}
\label{sec:intro}
When investigating higher-order behavior within a set of variables we are often interested in which variables carry information about others and how that information is distributed among them. For several variables carrying information about a chosen target variable, that information can be redundant (available from each of them), unique (available from one only) or synergistic (available only from their combination). Partial information decomposition (PID)~\cite{WilliamsBeer2010} makes this distinction quantitative and has been used to describe neural codes, environmental variables, machine-learning features and questionnaire items~\cite{Timme2014,Lizier2018,GoodwellKumar2017,Wollstadt2023,Marinazzo2024,VanSanten2026}. For two sources the joint mutual information splits into four atoms, but Shannon information theory supplies only three equations for them, so every bivariate PID is fixed by one further number, usually a definition of redundancy~\cite{WilliamsBeer2010,Bertschinger2013,Bertschinger2014,BanerjeeGriffith2015,Pica2017}. The many proposals~\cite{Harder2013,Bertschinger2014,GriffithKoch2014,Griffith2014GK,Barrett2015,Ince2017ccs,FinnLizier2018,James2019Idep,NiuQuinn2019,Makkeh2021,Kolchinsky2022,GoodwellKumar2017} rest on different intuitions about what it means for two sources to carry ``the same'' information and disagree on simple examples~\cite{Liardi2026}. This raises the important question of what each of these measures actually reports and which claims about the variables it supports. The sign of the co-information has long served as a coarse indicator of net redundancy or net synergy~\cite{McGill1954,SchneidmanBialekBerry2003,Angelini2010,Stramaglia2014,Timme2014}, and Goodwell and Kumar~\cite{GoodwellKumar2017} observed that its positive part is the smallest, and the smaller source--target information the largest, redundancy compatible with nonnegative atoms. For three or more sources, the Williams--Beer axioms, local positivity (nonnegative atoms) and the identity property cannot hold together~\cite{Rauh2014}.

Hi-Fi~\cite{OntiveroOrtega2025PRE} comes from a different direction. It is a feature-importance method: for a driver $X$ and a target $Y$ it conditions on subsets of the other variables and records how the dependence between $X$ and $Y$ changes. The largest decrease below the pairwise dependence is called redundancy, the largest increase synergy, and the minimum over conditioning sets the unique contribution of $X$. The construction originates in a min--max decomposition of the transfer entropy~\cite{Stramaglia2024PRR} and the original Hi-Fi measures dependence by leave-one-covariate-out predictability~\cite{OntiveroOrtega2025PRE}. Here, we study the information-theoretic version, which uses conditional mutual information~\cite{Lazic2026}. Several further extensions of Hi-Fi, e.g. pointwise~\cite{OntiveroOrtega2025local} and retraining-free~\cite{Aghilar2026} have followed. Hi-Fi uses the vocabulary of PID, but it is not built on the redundancy lattice, however, its equivalence to a two-source PID has already been suggested~\cite{Stramaglia2024PRR} and serves as a starting point for this work.

We answer the question raised above for PID and Hi-Fi by placing Hi-Fi on the segment of nonnegative bivariate PIDs and using the segment to compare measures. The answer sorts the claims these measures support into four levels, each requiring more than the one before. We parameterize the segment by the overlap $\delta=\min(\Red,\Syn)$, the common excess of redundancy and synergy over their net balance, which is the co-information. For two sources, Hi-Fi is the point $\delta=0$ of Goodwell and Kumar's segment, corresponding to a nonnegative PID whose redundancy and synergy are the positive and negative parts of the co-information, and which satisfies the usual axioms but neither the marginal assumption nor the Blackwell property of Bertschinger \emph{et al.} Its atoms bound those of every nonnegative PID, redundancy and synergy from below and the unique atoms from above. Hi-Fi's atoms, and with them these bounds, are functions of the three Shannon terms alone and are therefore identified (level~0), and the measures differ only in the overlap (Sec.~\ref{sec:two}). We show that Hi-Fi, a pairwise-reference PID defined below, the dependency-constraint measure $\Idep$, BROJA and minimum mutual information (MMI) can all be written as the largest co-information over nested families of reference distributions, which orders their overlaps on every distribution and makes each overlap a statement about the distribution relative to the family of the chosen measure (level~1; Sec.~\ref{sec:measures}). Splitting the co-information into the co-information of the pairwise maximum-entropy model and the third-order connected information~\cite{Amari2001,Schneidman2003} separates synergy generated by pairwise structure from synergy that needs a three-way term (Sec.~\ref{sec:triadic}). The triadic part can be tested against the pairwise model with the standard likelihood-ratio test (level~2), and the split defines a PID that adds overlap only when pairwise redundancy and triadic synergy coexist (Sec.~\ref{sec:pairref}). It furthermore shows how much of MMI's synergy, the largest of any nonnegative PID, is overlap (Sec.~\ref{sec:mmi}).

With three or more sources, Hi-Fi compares a driver with each set of partner sources, that is, it works on pairs of merged source groups. The bivariate results hold on every such pair, and MMI and Hi-Fi still give the smallest and the largest unique atoms that any nonnegative PID allows. Zero overlap, however, is often not attainable on all pairs at once. Two inequalities across pairs rule it out. In the case of three sources they are also sufficient, and for identity targets zero overlap is possible exactly when the sources' information diagram is nonnegative, which connects the obstruction to the counterexample of Rauh \emph{et al.}~\cite{Rauh2014} (Sec.~\ref{sec:multi}). Claims about mechanism, such as two sources acting jointly on the target, need assumptions from outside the distribution (level~3), and Section~\ref{sec:discussion} returns to the four levels to suggest reporting practices for Hi-Fi and PID results.

%% file: sec2_two.tex
\section{Two sources: Hi-Fi on the segment of nonnegative PIDs}
\label{sec:two}

\subsection{Partial information decomposition}
\label{sec:setup}

Let $Y$ be a target  and $X_1,X_2$ two sources with joint distribution $P$. Unless stated otherwise, the variables take finitely many values; jointly Gaussian variables are treated in Sec.~\ref{sec:gauss}. The mutual information, $I(Y;X_i)=H(Y) + H(X_i) - H(Y,X_i)$, is measured in bits and we write
\begin{equation}
\begin{aligned}
I_i &= I(Y;X_i), & I_{12}&=I(Y;X_1,X_2),\\
I_{i|j}&=I(Y;X_i\mid X_j), & i,j&\in\{1,2\},\ i\neq j.
\end{aligned}
\end{equation}
The co-information~\cite{McGill1954,Bell2003} of the triplet is
\begin{align}
C &= I_1+I_2-I_{12} = I_1-I_{1|2} \nonumber\\
  &= I(X_1;X_2)-I(X_1;X_2\mid Y).
\label{eq:coinfo}
\end{align}
It is symmetric in all three variables, with a positive value indicating net redundancy and a negative value net synergy. The interaction information~\cite{McGill1954} is the same quantity up to a sign convention, and for three variables the co-information's convention coincides with the O-information~\cite{Rosas2019}.

A PID of $P$ with two sources is a quadruple $(\Red,\Un_1,\Un_2,\Syn)$ that satisfies
\begin{subequations}
\label{eq:pid}
\begin{align}
I_1 &= \Un_1+\Red, \label{eq:pid1}\\
I_2 &= \Un_2+\Red, \label{eq:pid2}\\
I_{12} &= \Un_1+\Un_2+\Red+\Syn, \label{eq:pid3}
\end{align}
\end{subequations}
and it is nonnegative if all four atoms are. Every two-source PID satisfies
\begin{equation}
\Red-\Syn = C,\qquad I_{i|j} = \Un_i+\Syn ,
\label{eq:pidcons}
\end{equation}
the first relation going back to Williams and Beer~\cite{WilliamsBeer2010}. The three equations have the one-dimensional kernel $v=(+1,-1,-1,+1)$ in the coordinates $(\Red,\Un_1,\Un_2,\Syn)$, so a bivariate PID is fixed by any one of its atoms, usually $\Red$~\cite{WilliamsBeer2010,Bertschinger2013,Bertschinger2014,BanerjeeGriffith2015,Pica2017,ChicharroPanzeri2017}.

Measures of redundancy are selected and compared based on their properties. For two sources the commonly used ones are symmetry in the sources, self-redundancy $[\Red(X_1,X_1;Y)=I(Y;X_1)]$, monotonicity $[\Red\le\min(I_1,I_2)]$~\cite{WilliamsBeer2010}, and the identity property $[\Red(X_1,X_2;(X_1,X_2))=I(X_1;X_2)]$~\cite{Harder2013}. Bertschinger, Rauh, Olbrich, Jost, and Ay~\cite{Bertschinger2014} introduced the further assumption, referred to below as $\marg$, that the unique information of each source depends only on the source--target marginals $P(X_1,Y)$ and $P(X_2,Y)$. Since $I(Y;X_i)$ depends only on $P(X_i,Y)$, Eqs.~\eqref{eq:pid1} and~\eqref{eq:pid2} make $\marg$ equivalent to the same requirement on the redundancy~\cite[Lemma~2]{Bertschinger2014}, which is how Liardi \emph{et al.}~\cite{Liardi2026} states it. Their measure is known as BROJA~\cite{Bertschinger2014,GriffithKoch2014}, and its unique information vanishes if and only if the channel $P(x_i\mid y)$ is a garbling of $P(x_j\mid y)$, i.e.\ can be obtained from it through a channel that does not depend on $y$; taken as a requirement on a measure, this is the Blackwell property~\cite{Rauh2017}.

\subsection{Hi-Fi}
\label{sec:hifi}

Hi-Fi~\cite{OntiveroOrtega2025PRE,Lazic2026} decomposes the dependence between a target $Y$ and a driver $X$ with respect to an environment $\mathbf{Z}=\{Z_1,\dots,Z_m\}$ of other variables. In its information-theoretic form~\cite{Lazic2026}, the dependence given a subset $\mathbf W\subseteq\mathbf Z$ is $\Phi_{\mathbf W}=I(Y;X\mid\mathbf W)$, and the marginal dependence is $\Phi_\emptyset=I(Y;X)$. Hi-Fi defines
\begin{equation}
\begin{aligned}
U&=\min_{\mathbf W}\Phi_{\mathbf W},\qquad R=\Phi_\emptyset-U, \\
S&=\max_{\mathbf W}\Phi_{\mathbf W}-\Phi_\emptyset,
\end{aligned}
\label{eq:hifi}
\end{equation}
where the minimum and maximum run over all subsets $\mathbf W\subseteq\mathbf Z$ including $\mathbf W=\emptyset$, so that $U\le\Phi_\emptyset\le\max_{\mathbf W}\Phi_{\mathbf W}$, all three atoms are nonnegative, and $\max_{\mathbf W}\Phi_{\mathbf W}=U+R+S$~\cite{Stramaglia2024PRR,Lazic2026}. For a single driver, $U$ is the optimally conditioned transfer entropy of Ko\v{r}enek \emph{et al.}~\cite{Korenek2025}, who note that for two sources it yields a PID when the redundancy and synergy are obtained from the Shannon terms. Proposition~\ref{prop:hifipid} identifies this PID as Hi-Fi's.
In applications the subsets are found by a greedy forward search with surrogate-data testing at each step~\cite{OntiveroOrtega2025PRE,Lazic2026}. All results below are population statements about the information-theoretic version with exhaustive search, with greedy search covered by remark~\ref{rem:greedy}.

For a triplet the environment of the driver--target pair $(Y,X_i)$ is the single variable $X_j$, and Eq.~\eqref{eq:hifi} has the closed form
\begin{equation}
\begin{aligned}
U_{i}&=\min\left(I_i,\,I_{i|j}\right)=I_i-\pos{C},\\
R&=\pos{C},\qquad S=\pos{-C},
\end{aligned}
\label{eq:hifi3}
\end{equation}
with $\pos{C} = \max(0,C)$. $R$ and $S$ are the positive and negative parts of the signed co-information, so at most one of them is nonzero, and they are the same whichever variable is called the target. The two driver--target pairs $(Y,X_1)$ and $(Y,X_2)$ share $R$ and $S$ and differ only in $U_i$. Reading the sign of the co-information as net redundancy or net synergy is classical~\cite{McGill1954,SchneidmanBialekBerry2003,Timme2014} and underlies the Granger-causality decompositions from which Hi-Fi descends~\cite{Angelini2010,Stramaglia2014,Stramaglia2016}.

\subsection{The segment and the overlap}
\label{sec:overlap}

Goodwell and Kumar~\cite{GoodwellKumar2017} noted that a nonnegative bivariate PID must have $\pos{C}\le\Red\le\min(I_1,I_2)$ and called the two ends $R_{\min}$ and $R_{\mathrm{MMI}}$, the upper end being the minimum-mutual-information (MMI) redundancy used by Barrett~\cite{Barrett2015}. We restate their observation in the form used throughout.

\begin{proposition}[The segment of nonnegative PIDs]
\label{thm:segment}
For a fixed distribution $P(Y,X_1,X_2)$, every quadruple satisfying Eq.~\eqref{eq:pid} can be written as
\begin{equation}
\Red=\pos{C}+\delta,\quad \Syn=\pos{-C}+\delta,\quad \Un_i=U_i-\delta,
\label{eq:segment}
\end{equation}
with $\delta=\min(\Red,\Syn)\in\mathbb{R}$, and conversely Eq.~\eqref{eq:segment} satisfies Eq.~\eqref{eq:pid} for every $\delta\in\mathbb{R}$. The quadruple is nonnegative if and only if
\begin{equation}
\begin{aligned}
0\le&\delta\le\dmax, \; \text{with}\\
\dmax=\min\left(U_1,U_2\right)&=\min(I_1,I_2)-\pos{C}.
\end{aligned}
\label{eq:dmax}
\end{equation}
The point $\delta=0$ is the Hi-Fi PID $(\pos{C},U_1,U_2,\pos{-C})$ of Eq.~\eqref{eq:hifi3}, and the point $\delta=\dmax$ is MMI, $\Red_{\mathrm{MMI}}=\min(I_1,I_2)$.
\end{proposition}

\begin{proof}
Set $\delta=\Red-\pos{C}$. Then $\Syn=\Red-C=\pos{-C}+\delta$ by Eq.~\eqref{eq:pidcons}, and $\Un_i=I_i-\Red=U_i-\delta$ by Eqs.~\eqref{eq:pid1}--\eqref{eq:pid2} and~\eqref{eq:hifi3}; $\min(\Red,\Syn)=\delta$ because at least one of $\pos{C}$ and $\pos{-C}$ is zero. Conversely, Eq.~\eqref{eq:segment} is the quadruple $(\pos{C},U_1,U_2,\pos{-C})$ plus $\delta v$, and that quadruple satisfies Eq.~\eqref{eq:pid}: $\Un_i+\Red=I_i$ by Eq.~\eqref{eq:hifi3}, and $U_1+U_2+\pos{C}+\pos{-C}=I_1+I_2-\pos{C}+\pos{-C}=I_1+I_2-C=I_{12}$. Nonnegativity: $\Red,\Syn\ge0$ if and only if $\delta\ge0$, and $\Un_i\ge0$ if and only if $\delta\le\min(U_1,U_2)=\dmax$, which is nonnegative because $U_i=\min(I_i,I_{i|j})\ge0$. The end points follow by inserting the two redundancies.
\end{proof}

We call $\delta=\min(\Red,\Syn)$ the overlap of the PID, the common excess of redundancy and synergy over their net balance $\Red-\Syn=C$ (each of them exceeds its Hi-Fi value by $\delta$), and $\lambda=\delta/\dmax\in[0,1]$ its normalized position on the segment. Along the segment the Shannon terms, $C$ and $\Un_1-\Un_2=I_1-I_2$ are fixed, and $\delta$ bits leave each unique atom and enter both interaction atoms, so $\Red+\Syn=|C|+2\delta$. The identity $U_i=\Un_i+\delta$ holds for signed PIDs as well with Eq.~\eqref{eq:segment} showing which atom turns negative off the segment: for $\delta<0$ the smaller of Red and Syn, for $\delta>\dmax$ the unique atom of the weaker source. The signed measures $\Iccs$~\cite{Ince2017ccs}, $\Ipm$~\cite{FinnLizier2018} and $\Isx$~\cite{Makkeh2021} leave the segment often (Sec.~\ref{sec:numerics}). Goodwell and Kumar's rescaled redundancy is the segment parameterized by a normalized source coupling (Appendix~\ref{app:further}).

Table~\ref{tab:notation} lists where the symbols used for two sources are defined.

\begin{table}[tb]
\caption{An overview of the symbols used for two sources.}
\label{tab:notation}
\begin{ruledtabular}
\begin{tabular}{@{}lp{0.42\columnwidth}l@{}}
Symbol & Meaning & Defined in \\
\hline
$I_i,\ I_{12},\ I_{i|j}$ & Shannon terms & Sec.~\ref{sec:setup} \\
$C$ & co-information & Eq.~\eqref{eq:coinfo} \\
$\Red,\ \Un_i,\ \Syn$ & atoms of a PID & Eq.~\eqref{eq:pid} \\
$U_i,\ R,\ S$ & Hi-Fi's atoms & Eqs.~\eqref{eq:hifi}, \eqref{eq:hifi3} \\
$\delta,\ \dmax$ & overlap and its largest nonnegative value & Eqs.~\eqref{eq:segment}, \eqref{eq:dmax} \\
$\lambda$ & position on the segment & Sec.~\ref{sec:overlap} \\
$\Fam,\ \Red_\Fam,\ \delta_\Fam$ & reference family, its redundancy and overlap & Def.~\ref{def:family} \\
$\Ptwo,\ \Itwo$ & pairwise model and its co-information & Prop.~\ref{prop:split} \\
$\Ithree$ & triadic part of the co-information & Sec.~\ref{sec:triadic} \\
\end{tabular}
\end{ruledtabular}
\end{table}

\begin{proposition}[Properties of the Hi-Fi redundancy]
\label{prop:hifipid}
The quadruple $(\pos{C},U_1,U_2,\pos{-C})$ is a nonnegative PID (Proposition~\ref{thm:segment} with $\delta=0$). As a measure, $\Red_{\mathrm{HF}}=\pos{C}$ is symmetric, satisfies self-redundancy, monotonicity and the identity property, and vanishes whenever the sources are independent. It satisfies neither $\marg$ nor the Blackwell property.
\end{proposition}

\begin{proof}
Symmetry holds because $C$ is symmetric. If $X_2=X_1$, then $I_{12}=I_1$ and $C=I_1$. Monotonicity is satisfied through $C\le I_i$. If $Y=(X_1,X_2)$, then $I_i=H(X_i)$ and $I_{12}=H(X_1,X_2)$, so $C=I(X_1;X_2)\ge0$. If $X_1$ and $X_2$ are independent, the last form of Eq.~\eqref{eq:coinfo} gives $C=-I(X_1;X_2\mid Y)\le0$. For the last claim, let $Y$ be a uniform bit and the sources copies of $Y$ through binary symmetric channels with crossover $0.1$, either one shared copy ($X_1=X_2$) or two copies with independent noise: the source--target marginals agree ($I_1=I_2=0.531$ bits), but $C=0.531$ and $0.320$ bits respectively, so $U_i$ is $0$ in the first case and $0.211$ bits in the second, which violates $\marg$. In the second case the two channels $P(x_i\mid y)$ are identical, so each is a garbling of the other, yet $U_i=0.211$ bits, which violates the Blackwell property.
\end{proof}

Hi-Fi is distinguished by its position on the segment, namely, it has the smallest redundancy and synergy of any nonnegative PID, and also the largest unique atoms.

\begin{corollary}[Collapse and bounds]
\label{cor:bounds}
Let $(U_1,U_2,R,S)$ be the Hi-Fi atoms of a triplet. Every nonnegative PID of the triplet satisfies
\begin{equation}
\begin{aligned}
R\le\Red&\le R+\dmax,\quad S\le\Syn\le S+\dmax, \\
U_i-\dmax&\le\Un_i\le U_i,
\end{aligned}
\label{eq:bounds}
\end{equation}
and every bound is attained, at Hi-Fi or at MMI. In particular $S>0$ implies $\Syn>0$ and $R>0$ implies $\Red>0$ under every nonnegative PID, and if $I_i>I_j$ the stronger source has $\Un_i\ge I_i-I_j>0$ under every nonnegative PID, whereas unique information of the weaker source is a choice of measure. Moreover $\dmax=\min(I_1,I_{1|2},I_2,I_{2|1})$, so $\dmax=0$, and every nonnegative PID equals Hi-Fi, if and only if one source carries no information about the target ($I_j=0$) or none beyond the other source ($I_{j|i}=0$, e.g. the Markov chain $Y-X_i-X_j$).
\end{corollary}

\begin{proof}
Insert $0\le\delta\le\dmax$ into Eq.~\eqref{eq:segment}; $\Un_i\ge U_i-U_j=I_i-I_j$ because $\pos{C}$ cancels. The expression for $\dmax$ is Eq.~\eqref{eq:dmax} with Eq.~\eqref{eq:hifi3}, and all four terms are (conditional) mutual informations.
\end{proof}

The suppressor ($I_2=0$), XOR ($I_1=I_2=0$), and the redundant gate $X_1=X_2=Y$ are cases where the segment collapses to a point. Away from them the choice of measure matters. For the AND gate with uniform inputs, for example, $I_1=I_2=0.311$ and $I_{12}=0.811$ bits, so $C=-0.189$ and $\dmax=0.311$ bits, and the measures of Sec.~\ref{sec:measures} range from $\Red=0$ (Hi-Fi) to $0.311$ bits (BROJA, $\Imin$, MMI), with $\Idep$ at $0.082$ bits.

The choice of $\delta$ is a statement about what the unique information contains. $U_i=\min(I_i,I_{i|j})$ is the information about $Y$ that $X_i$ provides both on its own and on top of $X_j$. A PID with $\delta>0$ moves $\delta$ bits of it out of the unique atoms into redundancy and, since $\Red-\Syn$ is fixed, the same amount into synergy as well. Whether that is justified or appropriate depends on the distribution. In the AND gate nothing in $(I_1,I_2,I_{12})$ says whether the 0.311 bits that each input carries are the same content, so both $\delta=0$ and $\delta=\dmax$ are defensible, whereas in the triadic distribution of Sec.~\ref{sec:pairref} the shared bit and the parity bit are different bits, and $\delta=1$ bit is the natural choice. On the other hand, a PID with $\delta=0$ asserts that all of $U_i$ is unique to $X_i$, even when, as in the Blackwell counterexample of Proposition~\ref{prop:hifipid}, the other source carries a channel from which $X_i$ could be simulated. Nothing in the Shannon terms distinguishes that case either, so choosing between $\delta=0$ and $\delta>0$ needs information beyond the three Shannon terms, such as the specific information per target outcome ($\Imin$)~\cite{WilliamsBeer2010}, the distributions with the same source--target channels (BROJA)~\cite{Bertschinger2014}, or maximum-entropy reference models ($\Idep$)~\cite{James2019Idep}. Hi-Fi and MMI are the two extreme choices that use only the Shannon terms, MMI with the convention that the weaker source's information is fully contained in the stronger's. Which end is the conservative one depends on the claim: $\delta=0$ never asserts redundancy or synergy beyond the net balance, and is in that sense conservative for claims of synergy, but it asserts the largest unique information any nonnegative PID allows while $\delta=\dmax$ is the reverse. In the terms of the Introduction, the bounds of Corollary~\ref{cor:bounds} are level-0 statements, shared by every nonnegative PID, while the choice of $\delta$ is a level-1 statement; Section~\ref{sec:discussion} returns to which choice is conservative.

%% file: sec3_measures.tex
\section{Ordering the established measures}
\label{sec:measures}

\subsection{Reference families}

Several established measures can be written in a common form: each redundancy is the largest positive part of the co-information over some set of reference distributions, and the sets are nested. For $\Idep$ and MMI this is a derived representation rather than their defining construction (Appendix~\ref{app:idep}; the MMI set leaves the alphabet of $P$), but it is all the ordering below needs.

\begin{definition}[Reference-family redundancy]
\label{def:family}
Let $\Fam$ be a set of joint distributions of $(Y,X_1,X_2)$ on finite alphabets, such that $P\in\Fam$ and $I_Q(Y;X_i)=I_i$ for all $Q\in\Fam$ and $i=1,2$; the alphabets need not be those of $P$, a freedom used only by the MMI family below. Define
\begin{equation}
\Red_\Fam=\sup_{Q\in\Fam}\pos{C_Q},
\label{eq:redfam}
\end{equation}
where $C_Q$ is the co-information of $Q$, and complete it to a quadruple by Eq.~\eqref{eq:pid}.
\end{definition}
Each $Q\in\Fam$ is a reference distribution that shares with $P$ the features that define $\Fam$; only the co-informations $C_Q$ enter Eq.~\eqref{eq:redfam}, and the other atoms are constructed from the Shannon terms of $P$.

\begin{proposition}
\label{prop:family}
(a) $\Un_i^\Fam=\inf_{Q\in\Fam}U_i(Q)$: the unique information is the Hi-Fi unique information of the least favorable reference distribution, and the same $Q$ is least favorable for both sources. (b) $\Red_\Fam$ defines a nonnegative PID of $P$ with overlap $0\le\delta_\Fam\le\dmax$; it satisfies self-redundancy and monotonicity, and it is symmetric if exchanging the sources maps the family of $P$ onto the family of the exchanged $P$, as it does for every family below. (c) If $\Fam\subseteq\Fam'$, then $\delta_\Fam\le\delta_{\Fam'}$.
\end{proposition}

\begin{proof}
(a) For $Q\in\Fam$, $U_i(Q)=I_Q(Y;X_i)-\pos{C_Q}=I_i-\pos{C_Q}$, so taking the supremum of $\pos{C_Q}$ is taking the infimum of $U_i(Q)$, simultaneously for $i=1,2$. (b) $P\in\Fam$ gives $\Red_\Fam\ge\pos{C_P}$, i.e.\ $\delta_\Fam\ge0$. For every $Q\in\Fam$, $C_Q=I_Q(Y;X_i)-I_Q(Y;X_i\mid X_j)\le I_i$, so $\Red_\Fam\le\min(I_1,I_2)$, i.e.\ $\delta_\Fam\le\dmax$; Proposition~\ref{thm:segment} then gives a nonnegative PID, and monotonicity is the same bound. If $X_2=X_1$ in $P$, then $\pos{C_P}=I_1=\min(I_1,I_2)$ and both bounds coincide. (c) A supremum over a larger set is at least as large.
\end{proof}

\begin{figure}[tb]
\centering
\resizebox{\columnwidth}{!}{%
\begin{tikzpicture}[font=\footnotesize, >=Stealth]
  \definecolor{c1}{HTML}{2A78D6}
  \draw[rounded corners=7pt, fill=c1!4, draw=black!55] (0,0) rectangle (8.3,5.05);
  \node[anchor=north west, align=left] at (0.12,4.97) {$\QMI$: same $I(Y;X_1)$, $I(Y;X_2)$ $\Rightarrow$ MMI ($\delta=\dmax$)};
  \draw[rounded corners=6pt, fill=c1!8, draw=black!55] (0.3,0.25) rectangle (8.0,4.45);
  \node[anchor=north west, align=left] at (0.42,4.37) {$\DeltaP$: same $P(X_1,Y)$, $P(X_2,Y)$ $\Rightarrow$ BROJA};
  \node[anchor=north west, align=left, text=black!70] at (0.42,4.0) {($\Imin$, $\Ired$ and every measure obeying $\marg$: $\delta\ge\delta_{\mathrm{BROJA}}$)};
  \draw[rounded corners=5pt, dashed, fill=c1!13, draw=black!60] (0.6,0.5) rectangle (7.7,3.45);
  \node[anchor=north west] at (0.72,3.37) {$\Ftwo(P)\cup\{\QCI\}$ $\Rightarrow$ $\Idep$};
  \draw[rounded corners=4pt, fill=c1!20, draw=black!60] (0.9,0.75) rectangle (5.4,2.85);
  \node[anchor=north west, align=left] at (1.02,2.77) {$\Ftwo(P)$: same pairwise marginals\\$\Rightarrow$ pairwise reference};
  \fill (1.55,1.35) circle (1.6pt);
  \node[anchor=west, align=left] at (1.68,1.35) {$P$ $\Rightarrow$ Hi-Fi ($\delta=0$)};
  \fill (6.35,1.6) circle (1.6pt);
  \node[anchor=north, align=center] at (6.35,1.5) {$\QCI$};
  \draw[->, black!60, line width=0.6pt] (0.2,-0.3) -- node[below, text=black!80] {larger family $\Rightarrow$ larger overlap $\delta$} (8.1,-0.3);
\end{tikzpicture}}
\caption{Nested families of reference distributions. Each measure is the largest positive part of the co-information, $\pos{C_Q}$, over its family, Eq.~\eqref{eq:redfam}. A family that keeps less of $P$ fixed allows a larger co-information and hence a larger overlap. $\QCI=P(y)P(x_1|y)P(x_2|y)$ is the distribution in which the sources are conditionally independent given the target. $\Imin$ and $\Ired$ are not known to be of reference-family form, but Lemma~\ref{lem:maximal} places them at or above BROJA.}
\label{fig:families}
\end{figure}
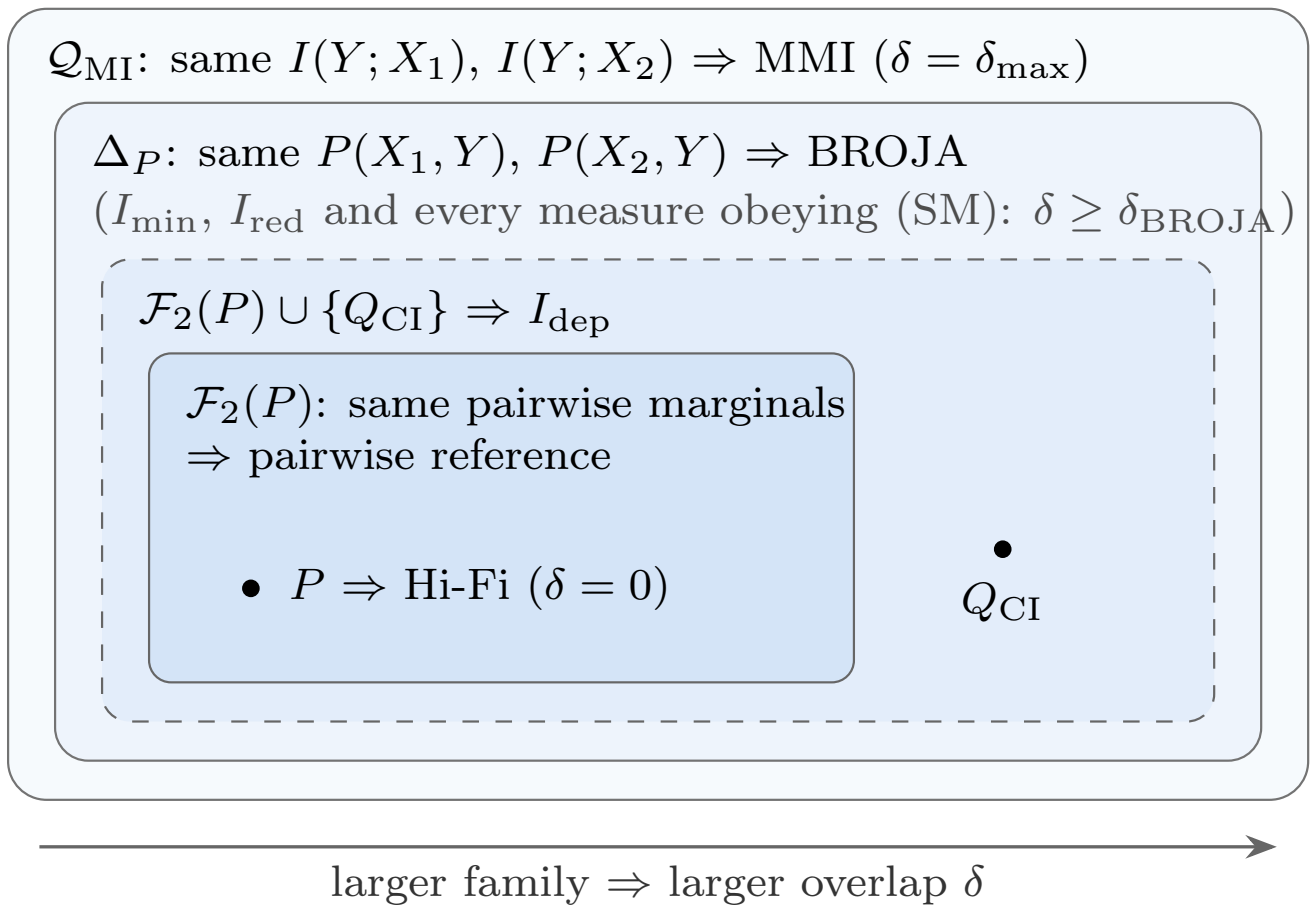

The named measures use three further families: $\Ftwo(P)$, the distributions on the alphabets of $P$ with the three bivariate marginals of $P$; $\DeltaP$, the distributions on the alphabets of $P$ with the source--target marginals $P(x_1,y)$ and $P(x_2,y)$~\cite{Bertschinger2014}; and $\QMI$, the distributions on arbitrary finite alphabets with $I_Q(Y;X_i)=I_i$ for $i=1,2$; one further reference distribution is $\QCI=P(y)P(x_1|y)P(x_2|y)$, in which the sources are conditionally independent given the target. Figure~\ref{fig:families} shows how they are nested.

\begin{proposition}[Named measures]
\label{prop:named}
Hi-Fi, the pairwise-reference PID of Sec.~\ref{sec:pairref}, $\Idep$~\cite{James2019Idep}, BROJA~\cite{Bertschinger2014} and MMI are the reference-family redundancies of $\{P\}$, $\Ftwo(P)$, $\Ftwo(P)\cup\{\QCI\}$, $\DeltaP$ and $\QMI$ respectively, with $\QCI=P(y)P(x_1|y)P(x_2|y)$, and their redundancies are $\pos{C}$, $\pos{\Itwo}$, $\max\{\Itwo,I_{\QCI}(X_1;X_2)\}$, $\max_{Q\in\DeltaP}C_Q$ and $\min(I_1,I_2)$, where $\Ptwo$ is the maximum-entropy distribution with the bivariate marginals of $P$ and $\Itwo$ its co-information. The families are nested,
\begin{equation}
\{P\}\subseteq\Ftwo(P)\subseteq\Ftwo(P)\cup\{\QCI\}\subseteq\DeltaP\subseteq\QMI .
\end{equation}
Hence, on every distribution,
\begin{equation}
0=\delta_{\mathrm{HF}}\le\delta_{(2)}\le\delta_{\Idep}\le\delta_{\mathrm{BROJA}}\le\delta_{\mathrm{MMI}}=\dmax .
\label{eq:order}
\end{equation}
\end{proposition}

\begin{proof}
(\emph{Pairwise reference}) For $Q\in\Ftwo(P)$ all one- and two-variable entropies equal those of $P$. Writing the co-information as the alternating sum
\begin{equation}
C=H_1+H_2+H_Y-H_{12}-H_{1Y}-H_{2Y}+H_{12Y},
\label{eq:altsum}
\end{equation}
$C_Q$ equals a constant plus $H_Q(X_1,X_2,Y)$ and is maximized by the maximum-entropy element $\Ptwo$ of $\Ftwo(P)$, which exists and is unique~\cite{DarrochRatcliff1972}.

(\emph{$\Idep$}) $\Idep$ is defined by James \emph{et al.}~\cite{James2019Idep} on a lattice of marginal constraints; the reference-family form is a derived representation, valid for two sources, obtained in Appendix~\ref{app:idep} from the four-edge minimum of Kay and Ince~\cite{KayInce2018}: the closed form $\max\{\Itwo,I_{\QCI}(X_1;X_2)\}$ equals $\max_{Q\in\Ftwo\cup\{\QCI\}}\pos{C_Q}$ because $C_{\QCI}=I_{\QCI}(X_1;X_2)-I_{\QCI}(X_1;X_2\mid Y)=I_{\QCI}(X_1;X_2)\ge0$. The second term was obtained independently by Lyu, Clark and Raviv~\cite{Lyu2025}, whose closed-form redundancy is $I(X_1';X_2)$ for a copy $X_1'$ of $X_1$ that is conditionally independent of $X_2$ given $Y$, that is, $C_{\QCI}$ alone. In the present terms it is $\sup$ over the one-element set $\{\QCI\}$, which does not contain $P$, so Proposition~\ref{prop:family}(b) does not apply: its synergy $C_{\QCI}-C$ is negative whenever $C_{\QCI}<C$, as for $X_2=X_1$ and $Y$ a noisy copy of $X_1$ (flip probability $0.1$: $C=0.531$, $C_{\QCI}=0.320$ bits).

(\emph{BROJA}) Bertschinger \emph{et al.}\ define the shared information as $\max_{Q\in\DeltaP}C_Q$~\cite{Bertschinger2014}, which is nonnegative because $\QCI\in\DeltaP$.

(\emph{MMI}) Let $X_a$ be the weaker source, $I_a\le I_b$. For every $Q\in\QMI$, $C_Q=I_Q(Y;X_a)-I_Q(Y;X_a\mid X_b)\le I_a$, with equality exactly when $Y-X_b-X_a$ is a Markov chain under $Q$. Such a chain need not exist in $\DeltaP$ (for $Y=(X_1,X_2)$ with independent uniform sources, $\DeltaP=\{P\}$), but it does in $\QMI$ when $I_b>0$: keep $P(X_b,Y)$ and let $X_a'$ equal $X_b$ with probability $1-\epsilon$ and an erasure symbol otherwise, independently of $(X_b,Y)$. Then $Y-X_b-X_a'$ is a chain and $I(Y;X_a')=(1-\epsilon)I_b$, which equals $I_a$ for $\epsilon=1-I_a/I_b$, so the bound $\min(I_1,I_2)$ is attained.

(\emph{Inclusions}) The first two are immediate; the source--target marginals are two of the three pairwise marginals and $\QCI$ has the source--target marginals of $P$, which gives the third; and $I_Q(Y;X_i)$ depends on $Q$ only through $Q(x_i,y)$, which gives the fourth. Each family thus satisfies Definition~\ref{def:family}, and Proposition~\ref{prop:family}(c) gives Eq.~\eqref{eq:order}.
\end{proof}

Each measure thus attributes to $P$ the net redundancy of the most redundant distribution that shares some of its features. Hi-Fi keeps everything; the pairwise reference removes three-way structure, which can only lower the co-information (Proposition~\ref{prop:split}); $\Idep$ adds the distribution in which the target is the only link between the sources, and calls at least the target-induced coupling $I_{\QCI}(X_1;X_2)$ redundant; BROJA keeps only the source--target channels; and MMI keeps only $I(Y;X_1)$ and $I(Y;X_2)$, so its extreme distribution is a chain in which the weaker source is a degraded copy of the stronger (Sec.~\ref{sec:mmi}). The overlap of a reference-family measure, $\delta_\Fam=\Red_\Fam-\pos{C}$, therefore measures how much more net redundancy the features that define $\Fam$ allow than $P$ itself has: a statement about $P$ relative to its family, at level~1 in the sense of the Introduction.

\subsection{Placing measures based on shared features}

A measure need not be of reference-family form to be placed on the segment: it is enough to know which features of $P$ its redundancy depends on. The argument is due to Bertschinger \emph{et al.}~\cite{Bertschinger2014}, who used it to show that $\Imin$ lies at or above BROJA; we state it in the present language.

\begin{lemma}[Bertschinger \emph{et al.}~\cite{Bertschinger2014}]
\label{lem:maximal}
Let a PID measure satisfy $\marg$ and be nonnegative on every distribution in $\DeltaP$, on the alphabets of $P$. Then its overlap at $P$ satisfies $\delta\ge\delta_{\mathrm{BROJA}}$.
\end{lemma}

\begin{proof}
Under $\marg$ the redundancy $\Red=I_1-\Un_1$ is the same for $P$ and for every $Q\in\DeltaP$. Nonnegative synergy for each such $Q$ requires $\Red\ge C_Q$, hence $\Red\ge\max_{Q\in\DeltaP}C_Q=\Red_{\mathrm{BROJA}}$.
\end{proof}

The argument uses only that $\DeltaP$ consists of all distributions sharing given features with $P$ and that the measure depends on $P$ only through those features, so it applies to the other families as well.

\begin{corollary}[Measures fixed by the features of a family]
\label{cor:features}
Let a PID measure be nonnegative on every distribution on finite alphabets.
\begin{enumerate}
\item[(i)] If its redundancy depends on $P$ only through the three bivariate marginals, then $\delta\ge\delta_{(2)}$. The pairwise-reference PID is the measure of this kind with the smallest overlap.
\item[(ii)] If its redundancy depends on $P$ only through $I(Y;X_1)$ and $I(Y;X_2)$, then $\Red=\min(I_1,I_2)=\Red_{\mathrm{MMI}}$ on every distribution.
\end{enumerate}
\end{corollary}

\begin{proof}
In both cases the redundancy takes the same value at $P$ and at every distribution $Q$ that shares the features, and nonnegative synergy at $Q$ requires $\Red\ge C_Q$, as in the proof of Lemma~\ref{lem:maximal}. (i) Taking $Q=\Ptwo$ gives $\Red\ge\Itwo$, and together with $\Red\ge0$, this gives $\Red\ge\pos{\Itwo}=\Red_{(2)}$, the pairwise-reference redundancy, which itself depends only on the bivariate marginals (Proposition~\ref{prop:split}). (ii) The erasure construction in the proof of Proposition~\ref{prop:named} gives a distribution with the same $I(Y;X_1)$ and $I(Y;X_2)$ as $P$ and co-information $\min(I_1,I_2)$, so $\Red\ge\min(I_1,I_2)$, and nonnegative unique atoms at $P$ give $\Red\le\min(I_1,I_2)$. If $I_1=I_2=0$, both bounds vanish. MMI is thus the only nonnegative PID whose redundancy measure depends solely on the two source-target information terms.
\end{proof}

Part (ii) needs the measure on alphabets larger than those of $P$, because the erasure construction adds a symbol; on the alphabets of $P$ the extreme chain need not exist. Both $\Imin$, whose nonnegativity for two sources is proved in Ref.~\cite{WilliamsBeer2010}, and Harder's $\Ired$, proved nonnegative in Ref.~\cite{Harder2013}, depend only on the source--target marginals, so Lemma~\ref{lem:maximal} places them at or above BROJA, and monotonicity caps both by $\min(I_1,I_2)$. Hence
\begin{equation}
\delta_{\mathrm{BROJA}}\le\delta_{\Imin},\,\delta_{\Ired}\le\dmax.
\label{eq:imin}
\end{equation}

Guaranteed positions thus come from nonnegativity (on the segment), reference-family form (ordered among the families) and $\marg$ with nonnegativity (at or above BROJA). Appendix~\ref{app:further} places the auxiliary-variable redundancy $I_{\mathrm{RAV}}$ between Hi-Fi and BROJA and gives the explicit position of $\Irr$~\cite{GoodwellKumar2017}; $I_{\mathrm{IG}}$~\cite{NiuQuinn2019} has no guaranteed position and falls on both sides of BROJA (Appendix~\ref{app:further}).

\subsection{Gaussian sources and target}
\label{sec:gauss}

For jointly Gaussian variables with a univariate target, write $r_1=\mathrm{corr}(Y,X_1)$, $r_2=\mathrm{corr}(Y,X_2)$ and $r_{12}=\mathrm{corr}(X_1,X_2)$, and let $\Sigma$ be the $3\times3$ correlation matrix. The definitions above carry over with densities, with $\Ftwo(P)$ the densities with the bivariate marginals of $P$ and $\DeltaP$ those with its source--target marginals; because every quantity is invariant under rescaling of the individual variables we take them standardized. A bivariate Gaussian with correlation $\rho$ has mutual information $-\tfrac12\log_2(1-\rho^2)$, so $I_i=-\tfrac12\log_2(1-r_i^2)$ and $I(Y;X_1\mid X_2)=-\tfrac12\log_2(1-\rho^2_{Y1\cdot2})$, with the partial correlation $\rho_{Y1\cdot2}=(r_1-r_2r_{12})/\sqrt{(1-r_2^2)(1-r_{12}^2)}$. Since $(1-r_2^2)(1-r_{12}^2)-(r_1-r_2r_{12})^2=\det\Sigma$, the co-information $C=I_1-I(Y;X_1\mid X_2)$ is
\begin{equation}
C=\tfrac12\log_2\frac{\det\Sigma}{(1-r_1^2)(1-r_2^2)(1-r_{12}^2)}.
\label{eq:gaussC}
\end{equation}

\begin{proposition}[Gaussian triplets]
\label{prop:gauss}
(i) $\Ptwo=P$, so $\Ithree=0$ and the pairwise-reference PID equals Hi-Fi. (ii) BROJA, $\Imin$, $\Ired$ and MMI coincide. This is Barrett's result~\cite{Barrett2015}, who proved it for every nonnegative PID whose redundancy depends only on the source--target marginals; the proof below gives the BROJA case through the chain distribution of Proposition~\ref{prop:named} and the $\Imin$ case as a check. (iii) $\Red_{\Idep}=\max\{\pos{C},-\tfrac12\log_2(1-r_1^2r_2^2)\}$ (Kay and Ince~\cite{KayInce2018}).
\end{proposition}

\begin{proof}
(i) Every $Q\in\Ftwo(P)$ has the bivariate marginals, hence the covariance matrix, of $P$. Among all distributions with this covariance the Gaussian one, which is $P$, has the largest entropy~\cite{CoverThomasBook}. Since $P\in\Ftwo(P)$, $\Ptwo=P$, so $\Ithree=0$ and $\Itwo=C$.

(ii) Let $X_a$ be the weaker source, $|r_a|\le|r_b|$, and $r_b\neq0$ (otherwise all redundancies vanish). The Gaussian $Q^\ast$ with the source--target correlations of $P$ and source correlation $r_a/r_b$ has a valid correlation matrix, since $\det\Sigma^\ast=(1-r_a^2/r_b^2)(1-r_b^2)\ge0$, lies in $\DeltaP$, and makes the partial correlation of $Y$ and $X_a$ given $X_b$ vanish, so $Y-X_b-X_a$ is a Markov chain and $C_{Q^\ast}=I_a$; hence $\Red_{\mathrm{BROJA}}\ge I_a=\Red_{\mathrm{MMI}}$, and Eq.~\eqref{eq:order} gives equality. For $\Imin$, the specific information of a source with correlation $r$, $I(Y{=}y;X)=\tfrac{1}{2\ln2}\left[-\ln(1-r^2)-r^2+r^2y^2\right]$, is nondecreasing in $r^2$ at every $y$, so the minimum over the sources is attained by $X_a$ at every $y$ and $\Red_{\Imin}=I_a$; $\Ired$ follows from Lemma~\ref{lem:maximal} and Eq.~\eqref{eq:imin}.

(iii) By Proposition~\ref{prop:named} and (i), $\Red_{\Idep}=\max\{C,I_{\QCI}(X_1;X_2)\}$; under $\QCI$ the sources are conditionally independent noisy copies of $Y$ with $\mathrm{corr}(X_1,X_2)=r_1r_2$, so $I_{\QCI}(X_1;X_2)=-\tfrac12\log_2(1-r_1^2r_2^2)$.
\end{proof}

In Gaussian triplets the named measures therefore occupy three points of the segment: Hi-Fi and the pairwise reference at $\lambda=0$, $\Idep$ in between (equal to Hi-Fi whenever the sources are at least as coupled as the target alone would make them), and BROJA, $\Imin$, $\Ired$ and MMI at $\lambda=1$, as Kay and Ince~\cite{KayInce2018} showed for $\Idep$ versus MMI; the information-geometric Gaussian PID of Kay~\cite{Kay2024}, which is not of reference-family form, lies close to but apart from both. The same positions hold for continuous variables with a Gaussian copula, because an invertible transformation of each variable maps $\Ftwo(P)$ onto $\Ftwo$ of the transformed distribution and leaves all mutual informations unchanged; they do not hold for discretized Gaussian-copula data such as Likert items (Sec.~\ref{sec:discussion}).

\subsection{Positions in practice}
\label{sec:numerics}

Table~\ref{tab:gates} gives the positions on named distributions: on AND and SUM, BROJA, $\Imin$ and MMI lie at $\lambda=1$ while $\Idep$ has $\lambda=0.26$ and $0.38$; on COPY, BROJA and $\Idep$ coincide with Hi-Fi at $\lambda=0$, whereas $\Imin$ and MMI are at 1 and violate the identity property; on RdnXor, Hi-Fi reports no redundancy and no synergy, yet every other measure in the table reports one bit of each (Sec.~\ref{sec:triadic}); on PwUnq~\cite{FinnLizier2018}, where each source carries its own half bit, Hi-Fi and the pairwise reference report no interaction while $\Imin$, BROJA and MMI report half a bit of redundancy and of synergy.

On ensemble A of Appendix~\ref{app:numerics}, 800 random distributions, the ordering of Eqs.~\eqref{eq:order} and~\eqref{eq:imin} held without a single violation (tolerance $10^{-4}$ bits). BROJA's median position was 1.00 for binary and 0.63 for ternary triplets, and $\Idep$ was the only established nonnegative measure that typically sat near Hi-Fi (median $\lambda$ between 0.13 and 0.23). BROJA reported $\Red>0$ and $\Syn>0$ together in 91--100\% of the distributions (both atoms above $10^{-4}$ bits; other thresholds in Appendix~\ref{app:numerics}), and the signed measures produced a negative atom in 26--97\% of them, depending on measure and ensemble. All BROJA values were computed with an exact convex program (Appendix~\ref{app:numerics}).

\begin{table*}[t]
\caption{Named distributions (bits; inputs uniform and independent unless stated). Columns: Shannon terms, co-information $C$, its pairwise and triadic parts $\Itwo$ and $\Ithree$ (Sec.~\ref{sec:triadic}), the length $\dmax$ of the segment of nonnegative PIDs, redundancy/synergy of Hi-Fi, the pairwise-reference PID (PR, Sec.~\ref{sec:pairref}), $\Idep$, BROJA and MMI, and the positions $\lambda=\delta/\dmax$ of $\Imin$ and of the signed measures $\Iccs$ and $\Ipm$ ($\lambda$ undefined, --, when $\dmax=0$, where every nonnegative PID equals Hi-Fi; signed measures can lie outside $[0,1]$, e.g.\ $\lambda_{\Ipm}=2.70$ and $5.98$ on noisy AND and the common driver reflect negative unique atoms of the weaker source, $\Un_2=-0.25$ and $-0.49$ bits). COPY, corr.: $Y=(X_1,X_2)$ with $P(X_1=X_2)=0.8$. Chain: binary symmetric channels with crossover 0.1 and 0.2. Common driver: copies of $Y$ with crossover 0.15 and 0.25. Noisy AND: output flipped with probability 0.1. Noisy copies: independent copies of $Y$ with crossover 0.1. Dyadic, triadic: Ref.~\cite{JamesCrutchfield2017}. PwUnq: Ref.~\cite{FinnLizier2018}.}
\label{tab:gates}
\begin{ruledtabular}
\footnotesize
\setlength{\tabcolsep}{3pt}
\begin{tabular}{lccccccccccccccc}
 & $I_1$ & $I_2$ & $I_{12}$ & $C$ & $\Itwo$ & $\Ithree$ & $\dmax$ & Hi-Fi & PR & $\Idep$ & BROJA & MMI & $\lambda_{\Imin}$ & $\lambda_{\Iccs}$ & $\lambda_{\Ipm}$\\
\hline
\input{tables/tab_gates}
\end{tabular}
\end{ruledtabular}
\end{table*}

\begin{figure*}[t]
\centering
\includegraphics[width=\textwidth]{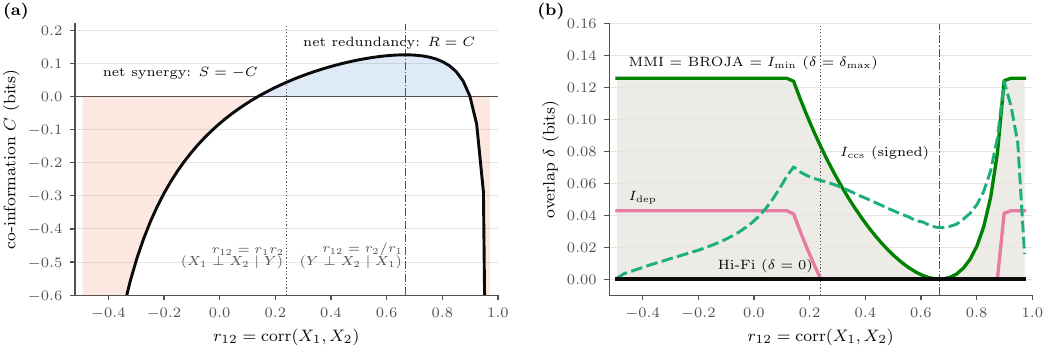}
\caption{Gaussian triplet with $\mathrm{corr}(Y,X_1)=0.6$ and $\mathrm{corr}(Y,X_2)=0.4$ as the source correlation $r_{12}$ varies. (a) The co-information $C=\Red-\Syn$ is the same for every PID; Hi-Fi reports $R=C$ where it is positive and $S=-C$ where it is negative. At $r_{12}=r_1r_2$ the sources are conditionally independent given $Y$. (b) Overlap $\delta$ chosen by each measure. The shaded band is the segment of nonnegative PIDs, from $\delta=0$ (Hi-Fi) to $\delta=\dmax$; MMI, BROJA and $\Imin$ coincide on its upper edge (Proposition~\ref{prop:gauss}). $\Iccs$ (Monte Carlo) leaves the band where one of its unique atoms is negative.}
\label{fig:gauss}
\end{figure*}

Figure~\ref{fig:gauss} shows a Gaussian triplet with $r_1=0.6$, $r_2=0.4$ as $r_{12}$ runs over its admissible range: net synergy appears for $r_{12}<0.14$ and $r_{12}>0.90$; MMI, BROJA and $\Imin$ trace the upper edge of the segment; $\Idep$ stays at the target-induced coupling (0.043 bits) wherever $C$ is smaller and joins Hi-Fi for $0.24\le r_{12}\le0.88$; and $\Iccs$ rises above the upper edge for $0.34\lesssim r_{12}\lesssim0.87$, where its unique atom for $X_2$ is negative. Over correlation matrices drawn uniformly from the admissible set (the Gaussian part of ensemble C, Appendix~\ref{app:numerics}), 62\% of Gaussian triplets are net synergistic. Appendix~\ref{app:tables} lists named Gaussian structures.

%% file: tables/tab_gates.tex
XOR & 0 & 0 & 1.000 & $-$1.000 & 0 & 1.000 & 0 & 0/1.00 & 0/1.00 & 0/1.00 & 0/1.00 & 0/1.00 & -- & -- & -- \\
AND & 0.311 & 0.311 & 0.811 & $-$0.189 & $-$0.189 & 0 & 0.311 & 0/0.19 & 0/0.19 & 0.08/0.27 & 0.31/0.50 & 0.31/0.50 & 1.00 & 0.33 & 1.80 \\
SUM & 0.500 & 0.500 & 1.500 & $-$0.500 & $-$0.500 & 0 & 0.500 & 0/0.50 & 0/0.50 & 0.19/0.69 & 0.50/1.00 & 0.50/1.00 & 1.00 & 0.00 & 1.00 \\
RDN & 1.000 & 1.000 & 1.000 & $+$1.000 & $+$1.000 & 0 & 0 & 1.00/0 & 1.00/0 & 1.00/0 & 1.00/0 & 1.00/0 & -- & -- & -- \\
UNQ & 1.000 & 0 & 1.000 & 0 & 0 & 0 & 0 & 0/0 & 0/0 & 0/0 & 0/0 & 0/0 & -- & -- & -- \\
COPY & 1.000 & 1.000 & 2.000 & 0 & 0 & 0 & 1.000 & 0/0 & 0/0 & 0/0 & 0/0 & 1.00/1.00 & 1.00 & 0.00 & 1.00 \\
COPY, corr. & 1.000 & 1.000 & 1.722 & $+$0.278 & $+$0.278 & 0 & 0.722 & 0.28/0 & 0.28/0 & 0.28/0 & 0.28/0 & 1.00/0.72 & 1.00 & 0.37 & 1.00 \\
RdnXor & 1.000 & 1.000 & 2.000 & 0 & $+$1.000 & 1.000 & 1.000 & 0/0 & 1.00/1.00 & 1.00/1.00 & 1.00/1.00 & 1.00/1.00 & 1.00 & 1.00 & 1.00 \\
chain $Y\!\to\!X_1\!\to\!X_2$ & 0.531 & 0.173 & 0.531 & $+$0.173 & $+$0.173 & 0 & 0 & 0.17/0 & 0.17/0 & 0.17/0 & 0.17/0 & 0.17/0 & -- & -- & -- \\
common driver & 0.390 & 0.189 & 0.489 & $+$0.090 & $+$0.090 & 0 & 0.098 & 0.09/0 & 0.09/0 & 0.09/0 & 0.19/0.10 & 0.19/0.10 & 1.00 & 1.89 & 5.98 \\
noisy AND & 0.147 & 0.147 & 0.412 & $-$0.119 & $-$0.041 & 0.077 & 0.147 & 0/0.12 & 0/0.12 & 0.03/0.15 & 0.15/0.27 & 0.15/0.27 & 1.00 & 0.44 & 2.70 \\
noisy copies & 0.531 & 0.531 & 0.742 & $+$0.320 & $+$0.320 & 0 & 0.211 & 0.32/0 & 0.32/0 & 0.32/0 & 0.53/0.21 & 0.53/0.21 & 1.00 & 1.22 & 2.35 \\
dyadic & 1.000 & 1.000 & 2.000 & 0 & 0 & 0 & 1.000 & 0/0 & 0/0 & 0/0 & 0/0 & 1.00/1.00 & 1.00 & 0.00 & 1.00 \\
triadic & 1.000 & 1.000 & 2.000 & 0 & $+$1.000 & 1.000 & 1.000 & 0/0 & 1.00/1.00 & 1.00/1.00 & 1.00/1.00 & 1.00/1.00 & 1.00 & 1.00 & 1.00 \\
PwUnq & 0.500 & 0.500 & 1.000 & 0 & 0 & 0 & 0.500 & 0/0 & 0/0 & 0.25/0.25 & 0.50/0.50 & 0.50/0.50 & 1.00 & 0.00 & 0.00 \\

%% file: sec4_triadic.tex
\section{Pairwise and triadic parts of the co-information}
\label{sec:triadic}

\subsection{The split}

Hi-Fi sees only the net co-information. The full distribution allows that net to be split into a part fixed by the pairwise marginals and a part that requires a three-way term. Let $\Ptwo$ be the maximum-entropy distribution with the three bivariate marginals of $P$ (equivalently, the maximum-likelihood fit of the pairwise log-linear model to $P$, computable by iterative proportional fitting~\cite{DarrochRatcliff1972,Bishop1975}), and let
\begin{equation}
\Ithree=D(P\,\|\,\Ptwo)=H(\Ptwo)-H(P)\ge0
\end{equation}
be the third-order connected information~\cite{Amari2001,Schneidman2003}. It vanishes exactly when $P$ belongs to the closure of the family of distributions whose log-linear expansion has no three-way term. The following identity is the co-information form of a known split: Hern\'andez, Zanette and Samengo~\cite{Hernandez2015} decomposed the three-variable multi-information into a pairwise part, fixed by the three bivariate marginals, and a genuinely triadic part, which is $\Ithree$.

\begin{proposition}
\label{prop:split}
$C=\Itwo-\Ithree$, where $\Itwo=C_{\Ptwo}$ is the co-information of the pairwise model and depends on the bivariate marginals of $P$ only.
\end{proposition}

\begin{proof}
$\Ptwo$ has the bivariate marginals of $P$, hence also its univariate marginals, so in the alternating sum, Eq.~\eqref{eq:altsum}, every term except $H_{12Y}$ is the same for $P$ and $\Ptwo$. Subtracting the two sums gives $\Itwo-C=H(\Ptwo)-H(P)=\Ithree$. Since the maximum-entropy distribution with given marginals is unique~\cite{DarrochRatcliff1972}, $\Itwo$ depends on $P$ only through its bivariate marginals.
\end{proof}

Since $C$ is the sum of the three pairwise mutual informations minus the total correlation, and the total correlation splits as $\mathrm{TC}(P)=\mathrm{TC}(\Ptwo)+\Ithree$ in the hierarchy of Amari and Schneidman \emph{et al.}~\cite{Amari2001,Schneidman2003}, Proposition~\ref{prop:split} is that hierarchy read through the co-information, and it restates the pairwise--triadic split of Ref.~\cite{Hernandez2015} with the pairwise part written as $\Itwo$; equivalently, it splits the triadic O-information~\cite{Rosas2019} into a pairwise-maximum-entropy part and a connected part, which bears on whether synergy requires higher-order mechanisms~\cite{Robiglio2025,Caprioglio2026}. Four consequences follow:
\begin{enumerate}
\item Triadic structure induces synergy. $\Ithree\ge0$ can only lower the co-information, whereas pairwise structure can push it either way, since $\Itwo<0$ in collider-like and suppressor-like pairwise models. AND and SUM are entirely pairwise and still net synergistic (Table~\ref{tab:gates}).
\item Hi-Fi's atoms split accordingly, $S=\pos{\Ithree-\Itwo}$ and $R=\pos{\Itwo-\Ithree}$: Hi-Fi synergy is triadic synergy plus pairwise-generated synergy minus pairwise redundancy.
\item A vanishing co-information can hide both. RdnXor has $\Itwo=+1$ bit of pairwise redundancy and $\Ithree=1$ bit of triadic synergy, which cancel to $C=0$; Hi-Fi reports neither. This is why every measure with $\delta>0$ in Table~\ref{tab:gates} reports one bit of each.
\item All Gaussian synergy is pairwise. By Proposition~\ref{prop:gauss}(i), $\Ithree=0$ for Gaussian triplets, and the same holds for continuous distributions with a Gaussian copula, because an invertible transformation of each variable maps the family without three-way term onto itself, hence $\Ptwo$ onto the pairwise projection of the transformed density, and leaves the divergence $D(P\|\Ptwo)$ unchanged~\cite{CoverThomasBook}. Net synergy in such models is therefore suppression in the regression sense (Remark~\ref{cor:suppression}); a collider is one way to produce it, a negative correlation among latent factors another.
\end{enumerate}

\begin{remark}[Gaussian synergy is suppression]
\label{cor:suppression}
For a Gaussian triplet, $C=\tfrac12\log_2[(1-\rho^2_{Y1\cdot2})/(1-r_1^2)]$, so $C<0$ if and only if $|\rho_{Y1\cdot2}|>|r_1|$: conditioning on $X_2$ increases the magnitude of the association between $Y$ and $X_1$, which is the regression definition of a suppressor~\cite{Conger1974}, and by the symmetry of Eq.~\eqref{eq:gaussC} the same holds for every pair of the triplet. Writing $a,b,c$ for the three correlations, $abc<0$ suffices for $C<0$: an antibalanced correlation triangle is net synergistic, the three-variable case of the result of Caprioglio \emph{et al.}~\cite{Caprioglio2026} that Gaussian systems with antibalanced signed correlation graphs are synergy-dominated in the O-information, and an instance of the negative higher-order information of frustrated systems noted by Matsuda~\cite{Matsuda2000}.
\end{remark}

\begin{figure*}[t]
\centering
\includegraphics[width=\textwidth]{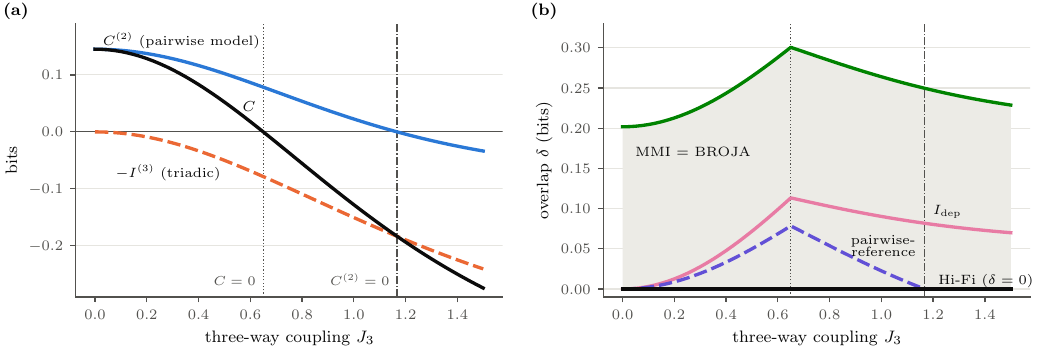}
\caption{Binary Ising triplet with $J_{1Y}=J_{2Y}=0.8$, $J_{12}=0$ and a three-way coupling $J_3$. (a) The co-information splits into the co-information of the pairwise maximum-entropy model, $\Itwo$, and the triadic connected information, $\Ithree$. Hi-Fi reports only their difference. (b) Overlap chosen by each measure; the shaded band is the segment of nonnegative PIDs, whose upper edge is MMI (BROJA coincides with it here). The pairwise-reference PID departs from Hi-Fi only while pairwise redundancy ($\Itwo>0$) and triadic synergy ($\Ithree>0$) coexist.}
\label{fig:ising}
\end{figure*}

An Ising triplet shows both parts at work. In the binary spin model
\begin{equation}
\begin{aligned}
p(s)\propto\exp\big(&J_{1Y}s_1s_Y+J_{2Y}s_2s_Y\\
&+J_{12}s_1s_2+J_3s_1s_2s_Y\big),\qquad s_i\in\{\pm1\},
\end{aligned}
\end{equation}
the pairwise couplings generate $\Itwo$ and the three-way coupling $J_3$ generates $\Ithree$. Only a term that is odd under the global spin flip $s\to-s$ can generate $\Ithree$: without external fields a flip-symmetric triplet has $p(s)=p(-s)$, so its log-linear expansion contains only even, i.e.\ pairwise, terms and $\Ithree=0$. This includes three-body couplings of the form $\delta_{s_1s_2s_3}=(1+s_1s_2+s_1s_3+s_2s_3)/4$, as in the simplicial Ising model of Robiglio \emph{et al.}~\cite{Robiglio2025}, and the Ising validation of the Gaussian results of Ref.~\cite{Caprioglio2026}: for an isolated triplet, or for the full-system distribution, the synergy of such models is pairwise-generated (the marginal of three spins embedded in a larger system can carry three-way terms, and dynamical variants of the O-information are not covered). Figure~\ref{fig:ising} adds a growing three-way coupling to a pairwise redundant structure. $C$ changes sign at $J_3\approx0.65$, where Hi-Fi's $R$ turns into $S$; $\Itwo$ also decreases, because the three-way term shifts the bivariate marginals, but changes sign only at $J_3\approx1.17$. Between the two crossings the pairwise structure is net redundant while the triplet as a whole is net synergistic; at $J_3=0.6$, $C=+0.018$ bits, so Hi-Fi reports almost nothing, while 0.087 bits of pairwise redundancy are offset by 0.069 bits of triadic synergy (Table~\ref{tab:ising} in Appendix~\ref{app:tables}).

\subsection{The pairwise-reference PID}
\label{sec:pairref}

Proposition~\ref{prop:split} turns into a decomposition when the pairwise model is used as the reference family. The pairwise-reference PID applies Hi-Fi to the pairwise model $\Ptwo$ and adds the triadic part $\Ithree$ to the synergy. Pairwise redundancy ($\Itwo>0$) and triadic synergy ($\Ithree>0$), which cancel in Hi-Fi's net co-information $C=\Itwo-\Ithree$, therefore stay apart. The PID equals Hi-Fi unless both are present, and then it departs from zero overlap by the smaller of the two (Proposition~\ref{prop:pairref}).

\begin{definition}[Pairwise-reference PID]
\label{def:pairref}
\begin{equation}
\begin{aligned}
\Red_{(2)}&=\pos{\Itwo},\qquad \Un_i=I_i-\pos{\Itwo},\\
\Syn_{(2)}&=\Ithree+\pos{-\Itwo}.
\end{aligned}
\label{eq:pairref}
\end{equation}
\end{definition}

The synergy has two parts: triadic synergy ($\Ithree$) and synergy generated by pairwise structure such as colliders and suppressors ($\pos{-\Itwo}$). Since $\Ptwo$ has the same $I_1$ and $I_2$ as $P$ and co-information $\Itwo$, the redundancy, the unique informations and the term $\pos{-\Itwo}$ are Hi-Fi's atoms on $\Ptwo$.

\begin{proposition}
\label{prop:pairref}
(i) Equation~\eqref{eq:pairref} is the reference-family PID of $\Ftwo(P)$, hence a nonnegative PID inside the segment, with overlap
\begin{equation}
\delta_{(2)}=\min\left\{\pos{\Itwo},\,\Ithree\right\}.
\label{eq:delta2}
\end{equation}
(ii) $\delta_{(2)}=0$ if and only if $\Ithree=0$ or $\Itwo\le0$. In particular the pairwise-reference PID equals Hi-Fi on every Gaussian triplet and on every strictly positive distribution without a three-way log-linear term.
(iii) $\Syn_{(2)}\ge\Ithree$, and $\delta_{(2)}$ is the smallest overlap of a nonnegative PID whose synergy is at least $\Ithree$. Consequently every measure with $\delta\ge\delta_{(2)}$, in particular $\Idep$, BROJA, $\Imin$ and MMI, reports at least $\Ithree$ bits of synergy.
(iv) It is symmetric in the sources, and its redundancy and synergy do not depend on which variable is called the target.
(v) It satisfies self-redundancy and the identity property.
(vi) It does not satisfy $\marg$, and $\delta_{(2)}\le\delta_{\mathrm{BROJA}}$.
\end{proposition}

\begin{proof}
(i) By Proposition~\ref{prop:named}, $\max_{Q\in\Ftwo}\pos{C_Q}=\pos{\Itwo}$, and Proposition~\ref{prop:family} applies. The synergy follows from $\Syn=\Red-C$ and $C=\Itwo-\Ithree$. For the overlap, $\delta_{(2)}=\pos{\Itwo}-\pos{\Itwo-\Ithree}$, which is 0 if $\Itwo\le0$ and $\min(\Itwo,\Ithree)$ otherwise. (ii) Follows from Eq.~\eqref{eq:delta2}, Proposition~\ref{prop:gauss}(i), and the fact that a strictly positive distribution without a three-way term is its own pairwise projection. (iii) $\Syn_{(2)}\ge\Ithree$ by definition. $\Syn=\pos{-C}+\delta$ increases with $\delta$, and since $\pos{-C}=\pos{\Ithree-\Itwo}$, $\Syn\ge\Ithree$ exactly when $\delta\ge\Ithree-\pos{\Ithree-\Itwo}$. The right-hand side equals $\delta_{(2)}$ if $\Itwo>0$ and is at most 0 otherwise. The measures listed have $\delta\ge\delta_{(2)}$ by Eqs.~\eqref{eq:order} and~\eqref{eq:imin}. (iv) $\Itwo$ and $\Ithree$ are symmetric in all three variables. (v) If $X_2=X_1$, or if $Y=(X_1,X_2)$, the bivariate marginals determine the joint, so $\Ftwo(P)=\{P\}$, $\Ptwo=P$, and the PID equals Hi-Fi, which has both properties (Proposition~\ref{prop:hifipid}). (vi) In the example in the proof of Proposition~\ref{prop:hifipid}, both distributions have $\Ithree=0$: the bivariate marginals determine the first, and the second, with conditionally independent sources, is a pairwise model. The PID therefore equals Hi-Fi on both, whose unique atoms differ although the source--target marginals agree. $\Ftwo(P)\subseteq\DeltaP$ gives the inequality.
\end{proof}

Table~\ref{tab:gates} compares the measure with others. Like $\Idep$ and BROJA, the pairwise-reference PID agrees with the intended decomposition on XOR, COPY, RdnXor, and the dyadic and triadic distributions. Unlike them, it adds no overlap where there is no triadic structure: on AND, noisy AND, noisy copies of the target and PwUnq it equals Hi-Fi, whereas BROJA adds 0.31, 0.15, 0.21 and 0.50 bits to both redundancy and synergy. On ensemble B of Appendix~\ref{app:numerics} (820 random distributions, distinct from ensemble A) it never produced an invalid atom, never left the segment, never violated the ordering Hi-Fi $\le$ pairwise reference $\le\Idep\le$ BROJA $\le$ MMI, and matched Eq.~\eqref{eq:delta2} to $5\times10^{-13}$ bits ($4\times10^{-6}$ bits on two sparse distributions, the tolerance of iterative fitting).

The pieces of Definition~\ref{def:pairref} are known, but the clipped decomposition itself appears not to have been proposed. Olbrich, Bertschinger and Rauh~\cite{Olbrich2015} showed that synergy measures derived from maximum-entropy projections onto interaction spaces are not compatible with nonnegative unique, shared and synergistic parts; the unclipped version, $\Red=\Itwo$ and $\Syn=\Ithree$, is implemented in \texttt{dit} as a reconstructability-analysis measure~\cite{Zwick2004} and gives $\Red=\Itwo=-0.19$ bits on AND, and Eq.~\eqref{eq:pairref} is its positive-part repair; Chicharro's maximum-entropy decompositions~\cite{Chicharro2017} and Ref.~\cite{Rosas2016} use related projections without the positive part. Related constructions use pairwise maximum-entropy references without a designated target~\cite{Rosas2016}, hierarchical families of channels~\cite{PerroneAy2016}, or maximum-entropy constraints on the source--target marginals only~\cite{Chicharro2017}; by Proposition~\ref{prop:named}, the pairwise-reference PID is $\Idep$ without the conditional-independence reference point, and $\Iccs$~\cite{Ince2017ccs} uses the same reference $\Ptwo$ but filters local co-information terms by sign, which makes it signed. On the interior of the simplex the maximum-entropy fit depends smoothly on the marginals, so $\Itwo$, $\Ithree$ and all atoms are continuous in $P$ there; continuity at the boundary and target monotonicity have not been examined.

The dyadic and triadic distributions of James and Crutchfield~\cite{JamesCrutchfield2017} show why some measure of this kind is needed. Both have $I_1=I_2=1$ and $I_{12}=2$ bits, so no PID whose redundancy is a function of $(I_1,I_2,I_{12})$ alone can tell them apart, yet in the dyadic distribution each pair of variables shares its own bit (one bit of unique information per source), whereas in the triadic one a bit is shared by all three variables and a parity bit links them (one bit of redundancy and one of synergy). The pairwise-reference PID separates them through $\Ithree$, which is 0 for the dyadic and 1 bit for the triadic distribution.

\paragraph*{Testing the triadic part.}
The plug-in estimate of $\Ithree$ is positive in every finite sample, so the triadic layer needs a test, and the appropriate one is the standard likelihood-ratio test of the log-linear model without three-way interaction~\cite{Bishop1975}: for $N$ observations, $G^2=2N\ln2\,\hat I^{(3)}$ is asymptotically $\chi^2$ with $(L_1-1)(L_2-1)(L_3-1)$ degrees of freedom for alphabet sizes $L_k$, under the usual condition that the expected cell counts of the fitted pairwise model are not small (a few counts or more in every cell~\cite{Bishop1975}); sparse tables call for exact or Monte Carlo $p$-values. Unlike an overlap, the claim $\Ithree>0$ can thus be tested against its model class, which is what places triadic synergy at level~2. The bias bookkeeping is worth recording: with $L$-level variables the plug-in co-information is biased by $-(L-1)^3/(2N\ln2)$ bits~\cite{Miller1955,Paninski2003}, and since Proposition~\ref{prop:split} also holds for plug-in estimates and the first-order biases of the bivariate tables cancel in $\hat C^{(2)}$, this bias sits, to leading order, entirely in $\hat I^{(3)}$. In a simulation at $N=2000$ the test held its nominal level on additive models and had full power on interaction models with $\Ithree\ge0.01$ bits (Appendix~\ref{app:numerics}).

%% file: sec6_mmi.tex
\section{MMI: the maximal overlap}
\label{sec:mmi}

MMI is widely used in neuroscience and physiology~\cite{Luppi2022,Varley2023} because it is simple and, for Gaussian systems with a univariate target, coincides with BROJA and $\Imin$~\cite{Barrett2015} (Proposition~\ref{prop:gauss}). Its maximal overlap gives it the largest synergy and the smallest unique atoms of any nonnegative PID, $\Syn_{\mathrm{MMI}}=\pos{-C}+\dmax$ (Corollary~\ref{cor:bounds}), an elementary consequence of monotonicity; since $\Red-\Syn=C$, MMI's synergy fractions are upper bounds among nonnegative PIDs rather than conservative estimates. With Proposition~\ref{prop:split}, MMI's synergy has two layers,
\begin{equation}
\Syn_{\mathrm{MMI}}=\underbrace{\dmax}_{\text{overlap}}+\underbrace{\pos{\Ithree-\Itwo}}_{\text{net synergy}=S},
\label{eq:layers}
\end{equation}
where the net synergy itself mixes pairwise-generated synergy ($\Itwo<0$) and triadic synergy ($\Ithree$). Only $\Ithree$ requires a three-way term in the distribution. The overlap $\dmax=\min(U_1,U_2)$ is a function of $P$, but the decision to place $\delta$ at $\dmax$ is a convention: it is the only part of Eq.~\eqref{eq:layers} that another nonnegative PID of the same distribution may set to zero.

Let $X_a$ be the weaker source, $I_a\le I_b$. Then $\Red_{\mathrm{MMI}}=I_a$, $\Un_a=0$ and, by Eq.~\eqref{eq:pidcons},
\begin{equation}
\Syn_{\mathrm{MMI}}=I(Y;X_a\mid X_b):
\label{eq:mmisyn}
\end{equation}
MMI labels the weaker source's marginal information as redundancy and its conditional information as synergy, as Barrett noted~\cite{Barrett2015}; Varley~\cite{Varley2024MMI} draws the dynamical consequence that MMI synergy can increase when a system is disintegrated, provided the first-order autocorrelations are kept. The redundancy $I_a$ is the co-information of a chain distribution $Q^\ast$ in which $X_a$ is a degraded copy of $X_b$ as far as $Y$ is concerned (Proposition~\ref{prop:named}), and MMI returns the same redundancy for $P$ and for $Q^\ast$, however differently the two sources are related. Such a $Q^\ast$ exists in $\DeltaP$ exactly when the channel $P(x_a\mid y)$ is a garbling of $P(x_b\mid y)$, which is when BROJA equals MMI~\cite{Bertschinger2014}, as for Gaussian triplets [Proposition~\ref{prop:gauss}(ii)]; for COPY it does not, and MMI calls one bit redundant although the two sources share nothing (Table~\ref{tab:gates}).

\paragraph*{How much of MMI's synergy is overlap.}
Table~\ref{tab:mmi} in Appendix~\ref{app:tables} quantifies Eq.~\eqref{eq:layers} on three simulated families: binary targets following an additive logistic model $P(Y=1\mid x_1,x_2)=\sigma(a+b_1x_1+b_2x_2)$ with a random input law, for which $\Ithree=0$ exactly (Proposition~\ref{prop:pairref}(ii)); the same model with an interaction term $cx_1x_2$; and Gaussian triplets with uniformly drawn correlation matrices. In the additive family, MMI reports synergy (more than 1\% of $I_{12}$) in 86\% of the triplets although the mechanism has no interaction on the logit scale; in 46\% MMI reports synergy although there is no net synergy at all, so that its synergy is entirely overlap; and over the family 68\% of the synergy MMI reports is $\dmax$. The remainder (Hi-Fi's $S$, 6.9\% of $I_{12}$ on average) is pairwise-generated. With an interaction term every measure rises, but 44\% of MMI's synergy is still overlap; there 71\% of the pairwise-reference synergy is triadic. In all three families the synergy fractions were ordered as Proposition~\ref{prop:named} predicts, without exception.

\paragraph*{MMI with a merged context.}
A PID of many variables is sometimes made tractable by merging them. For a focal variable $X_i$ and a target $Y$, all other variables $\mathbf Z$ are compressed into one context variable $E$, and a bivariate PID of $(X_i,E;Y)$ is reported for every pair. Equation~\eqref{eq:mmisyn} then has a direct consequence.

\begin{corollary}[MMI with a merged context]
\label{cor:merged}
If $I(Y;E)\ge I(Y;X_i)$, MMI on $(X_i,E;Y)$ gives
\begin{equation}
\Red=I(Y;X_i),\quad\Un_i=0,\quad\Syn=I(Y;X_i\mid E),
\label{eq:merged}
\end{equation}
and $\Un_E=I(Y;E)-I(Y;X_i)$. The premise holds in particular when $E$ keeps the information of $\mathbf Z$ about $Y$ and $\mathbf Z$ contains a variable at least as informative about $Y$ as $X_i$.
\end{corollary}

\begin{proof}
Eq.~\eqref{eq:mmisyn} with $X_a=X_i$ and $X_b=E$, and $I(Y;\mathbf Z)\ge I(Y;Z_j)$ for every $Z_j\in\mathbf Z$.
\end{proof}

Whether the premise holds is an empirical question about the embedding: a context that keeps most of the information of many variables will usually satisfy it, one compressed to a few categories need not. When it holds, the focal variable receives no unique information, the smallest value any nonnegative PID allows, whereas Hi-Fi's $U_i$ is the largest (for $E=\mathbf Z$ both statements are Theorem~\ref{thm:extremes} on the pair $(i,[n]\setminus i)$). Its conditional association with $Y$ given the context, including information that no other variable carries, is reported as synergy, which by Eq.~\eqref{eq:layers} is Hi-Fi's net synergy plus the overlap $\dmax=U_i$.

In the direct-edge example of Table~\ref{tab:gauss} ($X_1=0.8X_2+\varepsilon_1$, $Y=0.4X_1+0.8X_2+\varepsilon_2$, unit-variance noise) the mechanism is additive and the sources net redundant, yet MMI reports 0.107 bits of synergy, 17\% of $I_{12}$, which Hi-Fi assigns to the unique information of $X_1$.

A recent preprint on embedding-based PID of symptom networks~\cite{Hourican2026} combines a target-supervised context of this kind, built to retain the information of the other symptoms about the target, with MMI as the primary measure, and reports what Eq.~\eqref{eq:merged} predicts when the premise holds: ``the source-unique atom was zero or near-zero for all edges'' (its Sec.~3.2.2). It also reports that qualitative patterns ``were consistent across all five measures'' compared (MMI, $\Imin$, $\Irr$, $\Ipm$, $\IGK$) and suggests that MMI ``may underestimate synergy'' (its Sec.~2.5.1). The consistency across measures is guaranteed by construction as far as the redundancy- or synergy-dominance of a pair is concerned, since that call is the sign of the co-information of $(X_i,E;Y)$ under every PID [Eq.~\eqref{eq:pidcons}]; agreement in the ordering of edges by synergy fraction, which depends on $\delta$, is an empirical finding. The suggestion that MMI may underestimate synergy is the reverse of Corollary~\ref{cor:bounds}: among nonnegative PIDs, MMI's synergy is the largest (the signed $\Ipm$, which is in that comparison set, can exceed it, Table~\ref{tab:gates}). %

%% file: sec7_multi.tex
\section{Three or more sources}
\label{sec:multi}

Hi-Fi can be applied to many variables at once, so it matters which of the preceding results survive for sources $X_1,\dots,X_n$ with $n\ge3$. In brief: everything that holds for a triplet holds for every pair of merged source groups; MMI remains the extreme measure and Hi-Fi's unique atoms remain sharp bounds, attained by one nonnegative PID for all sources at once; but Hi-Fi itself is no longer a PID of the $n$ sources, and zero overlap on every pair it examines is often not attainable by any nonnegative PID. Throughout this section a \emph{nonnegative PID} means a nonnegative solution of the lattice equations for the one distribution at hand, with no axioms imposed across distributions.

\subsection{Lattice, free parameters, and coarse-grained pairs}

\begin{figure*}[t]
\centering
\input{figs/fig_lattice3}
\caption{The redundancy lattice for three sources. (a)~The 18 nodes, ordered upward by $\preceq$. Equation~\eqref{eq:shannon_n} fixes $I_\cap$ at the seven single-set nodes (gray); the six nodes $\{A\}\{B\}$ with disjoint $A,B$ (blue) are the coarse-grained pairs, where $I_\cap=\Red_{A,B}$; at the five dashed nodes nothing fixes $I_\cap$. (b,c)~The bivariate PID induced on two pairs by the assignment rule of the text: each region consists of the nodes at or below its bold node that lie in no region of a lower bold node. In (b) $X_3$ is ignored and the redundancy $\{1\}\{3\}$ counts as unique to $X_1$; in (c) every atom is kept and $\Un_{23}$ collects nine of them. (d)~The $q(3)=6$ pair nodes that Hi-Fi examines and fixes through $\delta_{i,Z}=0$.}
\label{fig:lattice3}
\end{figure*}
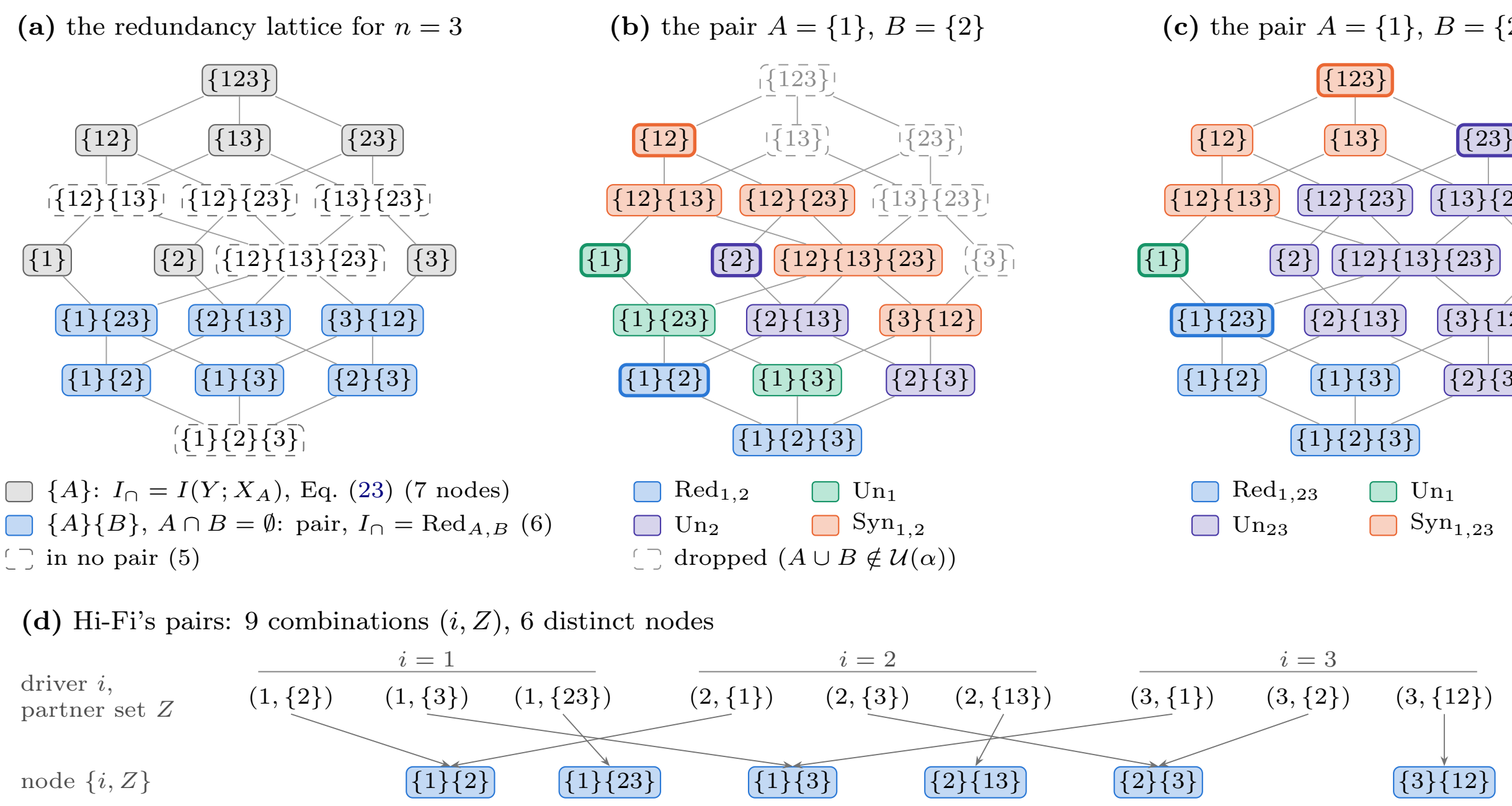

Let $[n]=\{1,\dots,n\}$ index the sources. The nodes of the redundancy lattice~\cite{WilliamsBeer2010,Gutknecht2021} are the antichains $\alpha$ of nonempty subsets of $[n]$. Write $\mathcal U(\alpha)=\{A\subseteq[n]:B\subseteq A\text{ for some }B\in\alpha\}$ for the up-set that $\alpha$ generates and order the nodes by $\alpha\preceq\beta$ if $\mathcal U(\beta)\subseteq\mathcal U(\alpha)$, that is, if every element of $\beta$ contains an element of $\alpha$; lower nodes have larger up-sets. A PID assigns an atom $\pi(\alpha)$ to every node such that
\begin{equation}
I(Y;X_A)=\sum_{\alpha:\,A\in\mathcal U(\alpha)}\pi(\alpha)
\label{eq:shannon_n}
\end{equation}
for every nonempty $A\subseteq[n]$, and its redundancy function is $I_\cap(\beta)=\sum_{\alpha\preceq\beta}\pi(\alpha)$, from which the atoms are recovered by M\"obius inversion. Since $A\in\mathcal U(\alpha)$ exactly when $\alpha\preceq\{A\}$, Eq.~\eqref{eq:shannon_n} states $I_\cap(\{A\})=I(Y;X_A)$. For $n=2$ it is Eq.~\eqref{eq:pid}; Fig.~\ref{fig:lattice3}(a) shows the 18 nodes for $n=3$.

The number of nodes is $D(n)-2$, where $D(n)$ is the Dedekind number, while Eq.~\eqref{eq:shannon_n} has $2^n-1$ independent equations: restricted to the single-set nodes $\{A'\}$ they form the zeta matrix of the Boolean lattice, which is triangular with unit diagonal, so given the atoms of the other nodes the single-set atoms follow uniquely. The PIDs of a distribution thus form an affine space of dimension $D(n)-2^n-1$, which is 1, 11, 151 and 7548 for $n=2,\dots,5$ (Table~\ref{tab:count}). The linear combinations of atoms that Eq.~\eqref{eq:shannon_n} fixes, i.e.\ on which every PID of $P$ agrees whatever the redundancy measure, are the Shannon invariants of Gutknecht \emph{et al.}~\cite{Gutknecht2025}; the net balances $C_{A,B}$ below are examples.

\begin{table}[tb]
\caption{Size of the problem for $n$ sources. Atoms: nodes of the redundancy lattice, $D(n)-2$. Shannon: equations~\eqref{eq:shannon_n}, $2^n-1$. Free: dimension of the space of PIDs. Pairs: unordered pairs of disjoint nonempty source groups, $(3^n-2^{n+1}+1)/2$. Hi-Fi pairs: those with a single source on one side, $q(n)=n(2^{n-1}-1)-\binom n2$. Unseen: free parameters (values of $I_\cap$) left in a PID that agrees with Hi-Fi on every pair it examines.}
\label{tab:count}
\begin{ruledtabular}
\begin{tabular}{rrrrrrr}
$n$ & Atoms & Shannon & Free & Pairs & Hi-Fi pairs & Unseen\\
\hline
\input{tables/tab_count}
\end{tabular}
\end{ruledtabular}
\end{table}

\paragraph*{Coarse-grained pairs.}
For disjoint nonempty $A,B\subseteq[n]$, merging the sources in $A$ and in $B$ gives a triplet $(Y;X_A,X_B)$ with co-information $C_{A,B}=I(Y;X_A)+I(Y;X_B)-I(Y;X_{A\cup B})$. Every $n$-source PID induces a bivariate PID of this triplet: the atom of node $\alpha$ goes to the redundancy if $A,B\in\mathcal U(\alpha)$, to the unique atom of $X_A$ if only $A\in\mathcal U(\alpha)$, to that of $X_B$ if only $B\in\mathcal U(\alpha)$, to the synergy if only $A\cup B\in\mathcal U(\alpha)$, and is dropped if $A\cup B\notin\mathcal U(\alpha)$ [Fig.~\ref{fig:lattice3}(b,c)]. By Eq.~\eqref{eq:shannon_n} the induced quadruple satisfies Eq.~\eqref{eq:pid} for the merged triplet, with redundancy $\Red_{A,B}=I_\cap(\{A,B\})$, because $\alpha\preceq\{A,B\}$ exactly when $A,B\in\mathcal U(\alpha)$. Proposition~\ref{thm:segment} therefore applies to every coarse-grained pair:
\begin{equation}
\begin{aligned}
\Red_{A,B}&=\pos{C_{A,B}}+\delta_{A,B},\\
\Syn_{A,B}&=\pos{-C_{A,B}}+\delta_{A,B}, 
\end{aligned}
\label{eq:coarse}
\end{equation}
and a nonnegative $n$-source PID has $0\le\delta_{A,B}\le\min(I_A,I_B)-\pos{C_{A,B}}$ for every pair, with $I_A=I(Y;X_A)$. The overlap $\delta_{A,B}=I_\cap(\{A,B\})-\pos{C_{A,B}}$ is linear in the atoms. Leaving the segment of a pair implies a negative atom, but for $n\ge3$ the converse fails, because the pairs constrain $I_\cap$ only at the nodes $\{A\}$ and $\{A,B\}$: for $Y=(X_1,X_2,X_3)$ with independent uniform bits, raising $I_\cap(\{1\}\{2\}\{3\})$ by $\epsilon$ above the PID with one unique bit per source keeps every pair at $\delta_{A,B}=0$ yet makes four atoms negative.

\paragraph*{Hi-Fi with $n$ sources.}
For a driver $X_i$ the environment is $X_{[n]\setminus i}$, and $\Phi_Z=I(Y;X_i\mid X_Z)=I_i-C_{i,Z}$, writing $i$ for $\{i\}$. Equation~\eqref{eq:hifi} becomes
\begin{equation}
\begin{aligned}
U_i&=I_i-\max_{Z}\pos{C_{i,Z}},\\
R_i=\max_{Z}&\pos{C_{i,Z}},\quad S_i=\max_{Z}\pos{-C_{i,Z}},
\end{aligned}
\label{eq:hifi_n}
\end{equation}
with $Z$ ranging over the nonempty subsets of $[n]\setminus\{i\}$ [Fig.~\ref{fig:lattice3}(d)]; $Z_R$ and $Z_S$ denote partner sets attaining $R_i$ and $S_i$. Each term is the bivariate Hi-Fi decomposition of the coarse-grained pair $(i,Z)$, i.e.\ the point $\delta_{i,Z}=0$ of its segment. Two differences from $n=2$ follow. First, $R_i$ and $S_i$ can both be positive, attained with different partners; in the random distributions of Table~\ref{tab:multi} both exceeded $\max(0.05\,I_i,10^{-3})$ bits for 11--67\% of the drivers, identity targets excepted, for which $C_{i,Z}=I(X_i;X_Z)\ge0$ and hence $S_i=0$. Second, Hi-Fi's numbers concern at most the nodes $\{A\}$ and $\{i,Z\}$, of which there are $2^n-1+q(n)$, and the $q(n)$ constraints $\delta_{i,Z}=0$ are independent of each other and of Eq.~\eqref{eq:shannon_n}, since each fixes $I_\cap$ at a distinct node. A PID that agreed with Hi-Fi on every pair it examines would still have 5 free parameters for three sources [the dashed nodes of Fig.~\ref{fig:lattice3}(a)] and 129 for four (Table~\ref{tab:count}).

\subsection{What carries over}

Both ends of the segment survive for $n$ sources (Theorem~\ref{thm:extremes}). Both results rest on one construction, which builds a PID from minima of a set function and gives its atoms in closed form. For two sources with $I_1\le I_2\le I_{12}$, cut $[0,I_{12})$ at these values into three layers: the bottom layer is reached by $X_1$ and $X_2$ and becomes redundancy, the middle one only by $X_2$ and becomes its unique information, the top one only by $(X_1,X_2)$ and becomes synergy; these are MMI's atoms. For $n$ sources each layer goes whole to the node formed by the minimal sets that reach it, and all nonzero atoms lie on one chain.

\begin{lemma}[Layer cake]
\label{lem:layercake}
Let $g$ be a set function on $[n]$ with $g(\emptyset)=0$ that is nondecreasing, $g(A)\le g(A')$ for $A\subseteq A'$. Define a redundancy function from the minima of $g$,
\begin{equation}
I_\cap^g(\beta)=\min_{B\in\beta}g(B),
\label{eq:layercake}
\end{equation}
and let $\pi^g$ be its atoms, $I_\cap^g(\beta)=\sum_{\alpha\preceq\beta}\pi^g(\alpha)$. Let $0=w_0<w_1<\dots<w_m=g([n])$ be the distinct values of $g$, and for $k=1,\dots,m$ let $\alpha_k$ be the node formed by the minimal sets $A$ with $g(A)\ge w_k$. Then
\begin{enumerate}
\item[(i)] $\alpha_1\prec\alpha_2\prec\dots\prec\alpha_m$, so these nodes form a chain;
\item[(ii)] $\pi^g(\alpha_k)=w_k-w_{k-1}$, and $\pi^g(\alpha)=0$ at every other node; in particular $\pi^g\ge0$;
\item[(iii)] $\sum_{\alpha:\,A\in\mathcal U(\alpha)}\pi^g(\alpha)=g(A)$ for every nonempty $A\subseteq[n]$, so $\pi^g$ satisfies Eq.~\eqref{eq:shannon_n} with $g(A)$ in place of $I(Y;X_A)$.
\end{enumerate}
(If $g\equiv0$, then $m=0$, $\pi^g\equiv0$, and the statements are vacuous.)
\end{lemma}

\begin{proof}
For $k=1,\dots,m$ let $\mathcal F_k=\{A:g(A)\ge w_k\}$, the sets that reach the $k$th layer $[w_{k-1},w_k)$. If $A\in\mathcal F_k$ and $A\subseteq A'$, then $g(A')\ge w_k$, so $\mathcal F_k$ is an up-set; it contains $[n]$ and not $\emptyset$, so its minimal elements form a node $\alpha_k$ with $\mathcal U(\alpha_k)=\mathcal F_k$.

(i) The sets with $g(A)=w_k$ lie in $\mathcal F_k$ but not in $\mathcal F_{k+1}$, so $\mathcal F_1\supsetneq\mathcal F_2\supsetneq\dots\supsetneq\mathcal F_m$, hence $\alpha_1\prec\dots\prec\alpha_m$ with strict inequalities.

(ii) Let the candidate $\pi$ put $w_k-w_{k-1}$ on $\alpha_k$ for each $k$ and $0$ elsewhere. For a node $\beta$, $\alpha_k\preceq\beta$ means $\mathcal U(\beta)\subseteq\mathcal F_k$, which, $\mathcal F_k$ being an up-set, holds exactly when every $B\in\beta$ lies in $\mathcal F_k$, i.e.\ when $w_k\le\min_{B\in\beta}g(B)$. The minimum is one of the values, say $w_j$, so $\alpha_k\preceq\beta$ exactly for $k\le j$, and $\sum_{\alpha\preceq\beta}\pi(\alpha)=\sum_{k\le j}(w_k-w_{k-1})=w_j=I_\cap^g(\beta)$. M\"obius inversion is unique, so $\pi=\pi^g$.

(iii) Since $A\in\mathcal U(\alpha)$ exactly when $\alpha\preceq\{A\}$, the sum equals $I_\cap^g(\{A\})=g(A)$.
\end{proof}

The name comes from the layer-cake representation of a nonnegative number as a stack of slices: for $x=I_\cap^g(\beta)$, the slice at height $t$ is present exactly when every $g(B)$ with $B\in\beta$ exceeds $t$, and the slices group into the layers of the lemma. The lemma is not new in substance: that a sum of minima of monotone set functions has nonnegative atoms is the structure of Williams and Beer's nonnegativity proof for $\Imin$~\cite{WilliamsBeer2010}, of Finn and Lizier's proof for their pointwise measures~\cite{FinnLizier2018}, and of the proof of Mages \emph{et al.}~\cite{Mages2024} for their pointwise Blackwell-specific measure; local positivity of MMI for $n\ge3$ is recorded in Ref.~\cite{Liardi2026}. What it adds is the explicit chain and the closed-form atoms, which Theorem~\ref{thm:extremes}(ii) uses. With $g(A)=I(Y;X_A)$, $I_\cap^g$ is MMI. More generally, if $I(Y;X_A)=\sum_kg_k(A)$ with every $g_k$ nondecreasing and $g_k(\emptyset)=0$, then $I_\cap(\beta)=\sum_k\min_{B\in\beta}g_k(B)$ is a nonnegative PID, and every nonnegative PID is of this form, with one piece $g_\alpha(A)=\pi(\alpha)\,\mathbf 1[A\in\mathcal U(\alpha)]$ per node. Nonnegative PIDs thus differ only in how they split $I(Y;X_A)$; MMI does not split it, and $\pi_{\mathrm{HF}}$ below splits off $U_i\,\mathbf 1[i\in A]$ for every $i$.

\begin{theorem}[Extremes for $n$ sources]
\label{thm:extremes}
Let $n\ge2$ and let $\mathcal P$ be the set of nonnegative PIDs of $P$.
\begin{enumerate}
\item[(i)] MMI, $I_\cap(\beta)=\min_{B\in\beta}I(Y;X_B)$, belongs to $\mathcal P$, and its redundancy function is the pointwise maximum over $\mathcal P$. Consequently MMI has the largest overlap on every coarse-grained pair at once, its synergy atom $\pi(\{[n]\})=\min_iI(Y;X_i\mid X_{[n]\setminus i})$ is the largest value of that atom in $\mathcal P$, and its unique atoms $\Un_i=\pos{I_i-I(Y;X_{[n]\setminus i})}$ are the smallest.
\item[(ii)] Every PID in $\mathcal P$ has $\Un_i\le U_i$, Hi-Fi's exhaustive unique atom, and a single PID $\pi_{\mathrm{HF}}\in\mathcal P$ attains $\Un_i=U_i$ for all $i$ at once: it places $U_i$ on the unique atoms and applies Lemma~\ref{lem:layercake} to $g(A)=I(Y;X_A)-\sum_{i\in A}U_i$. For $n=2$, $\pi_{\mathrm{HF}}$ is Hi-Fi.
\item[(iii)] Every PID in $\mathcal P$ has $\Red_{i,Z}\ge\pos{C_{i,Z}}$ and $\Syn_{i,Z}\ge\pos{-C_{i,Z}}$ for every driver $i$ and partner set $Z$. Hence $R_i>0$ implies redundancy between $X_i$ and $X_{Z_R}$ under every nonnegative PID, and $S_i>0$ implies synergy between $X_i$ and $X_{Z_S}$.
\end{enumerate}
\end{theorem}

\begin{proof}
(i) Membership is Lemma~\ref{lem:layercake} with $g(A)=I(Y;X_A)$. Pointwise maximality uses only nonnegativity: for $\pi\in\mathcal P$ and $B\in\beta$ we have $\beta\preceq\{B\}$, and nonnegative atoms make $I_\cap$ nondecreasing on the lattice, so $I_\cap(\beta)\le I(Y;X_B)$. The overlap statement follows from $\Red_{A,B}=I_\cap(\{A,B\})$. The atom of $\{[n]\}$ is counted in $I(Y;X_{[n]})$ and in no $I(Y;X_B)$ with $B\neq[n]$, and the other atoms counted in $I(Y;X_{[n]})-I(Y;X_{[n]\setminus i})$ are nonnegative, so it is at most this difference for every $i$; in the layer cake it receives the top layer, whose height is the smallest of these differences. On the pair $(i,[n]\setminus i)$ the only atom counted as unique to $X_i$ is $\pi(\{i\})$, so $\Un_i=I_i-\Red_{i,[n]\setminus i}\ge I_i-\min\{I_i,I(Y;X_{[n]\setminus i})\}$, with equality for MMI. (ii) The atom of $\{i\}$ is counted in $I(Y;X_{Z\cup i})$ but not in $I(Y;X_Z)$, and the other atoms counted in the difference are nonnegative, so $\pi(\{i\})\le I(Y;X_i\mid X_Z)$ for every $Z\subseteq[n]\setminus\{i\}$, including $Z=\emptyset$; hence $\Un_i\le U_i$. For $i\notin A$, $g(A\cup i)-g(A)=I(Y;X_i\mid X_A)-U_i\ge0$, so $g$ is nondecreasing, and $\pi_{\mathrm{HF}}$, the layer cake of $g$ plus $U_i$ on each node $\{i\}$, is the split $I(Y;X_A)=g(A)+\sum_iU_i\,\mathbf 1[i\in A]$ of the representation above, with the pieces $U_i\,\mathbf 1[i\in A]$ each contributing the layer cake $U_i$ on $\{i\}$; it is therefore nonnegative and satisfies Eq.~\eqref{eq:shannon_n}, hence lies in $\mathcal P$; by the first half its unique atoms are at most $U_i$, and by construction they are at least $U_i$. For $n=2$, $g(\{i\})=\pos{C}$ and $g(\{1,2\})=|C|$, so the layer cake puts $\pos{C}$ on redundancy and $\pos{-C}$ on synergy. (iii) Eq.~\eqref{eq:coarse} with $\delta_{i,Z}\ge0$, which holds because the induced bivariate atoms are sums of nonnegative atoms.
\end{proof}

Part (i) restates, in lattice form, that MMI is nonnegative and that the Williams--Beer monotonicity axiom caps every nonnegative redundancy by MMI's; parts (ii) and (iii) are the Hi-Fi side.

\begin{remark}[Greedy search]
\label{rem:greedy}
Parts (ii) and (iii) do not require an exhaustive search: every conditional information $\Phi_Z$ that a greedy search visits bounds $\Un_i$ from above and every visited $\pos{C_{i,Z}}$ bounds $\Red_{i,Z}$ from below, so a search that misses the best partner set returns valid but looser bounds. It can miss an entire bit: for the parity $Y=X_1\oplus X_2\oplus X_3$, exhaustive Hi-Fi gives $S_1=1$ bit and a greedy forward search gives $0$, because no single partner changes $I(Y;X_1)$; and for $Y=X_1$, $X_2=X_1\oplus N$, $X_3=N$ with $N$ a uniform bit, exhaustive Hi-Fi gives $R_1=1$ bit and $U_1=0$, the greedy search $R_1=0$ and $U_1=1$ bit.
\end{remark}

\subsection{What does not carry over}

For two sources, largest unique atoms, smallest redundancy and synergy, and zero overlap are one property (Proposition~\ref{thm:segment}). For $n$ sources the first survives (Theorem~\ref{thm:extremes}(ii)), but zero overlap on all pairs can fail, because pairs share atoms and constrain each other. Proposition~\ref{prop:acrosspairs} gives two such constraints: (M) can force a positive overlap on a single pair (Example~\ref{ex:copy}), and (S) can force the overlaps of three pairs to add up to a positive amount (Example~\ref{ex:rauh}). For three sources they are the only obstruction (Theorem~\ref{thm:three}).

\begin{proposition}[Constraints across pairs]
\label{prop:acrosspairs}
Let $A$, $B$, $B'$ be pairwise disjoint nonempty subsets of $[n]$. Every PID in $\mathcal P$ satisfies
\begin{align}
&\text{(M)}\quad \Red_{A,B}\le\Red_{A,B\cup B'},\\
&\text{(S)}\quad \Red_{A,B\cup B'}\le\Red_{A,B}+\Red_{A,B'}+\Syn_{B,B'}.
\end{align}
Consequently
\begin{align}
\delta_{A,B\cup B'}&\ge\pos{C_{A,B}}-\pos{C_{A,B\cup B'}},\label{eq:dM}\\
\delta_{A,B}+\delta_{A,B'}+\delta_{B,B'}&\ge\pos{C_{A,B\cup B'}}-\pos{C_{A,B}}\nonumber\\
&\quad-\pos{C_{A,B'}}-\pos{-C_{B,B'}},\label{eq:dS}
\end{align}
and if either right-hand side is positive, no nonnegative PID has zero overlap on all of these pairs.
\end{proposition}

\begin{proof}
The atom of a node $\alpha$ is counted in $\Red_{A,B}$ if $\alpha$ has an element inside $A$ and one inside $B$, and in $\Syn_{B,B'}$ if it has no element inside $B$ and none inside $B'$, but one inside $B\cup B'$. (M): an element inside $B$ is also inside $B\cup B'$, so every atom counted in $\Red_{A,B}$ is counted in $\Red_{A,B\cup B'}$; equivalently, $\{A,B\}\preceq\{A,B\cup B'\}$ and $I_\cap$ is nondecreasing. (S): let the atom of $\alpha$ be counted in $\Red_{A,B\cup B'}$, so that $\alpha$ has an element inside $A$ and an element $E$ inside $B\cup B'$. If some element of $\alpha$ lies inside $B$, the atom is counted in $\Red_{A,B}$; if some element lies inside $B'$, in $\Red_{A,B'}$; if neither, $E$ lies inside $B\cup B'$ while no element lies inside $B$ or $B'$, so it is counted in $\Syn_{B,B'}$. The right-hand side may count an atom twice and contains atoms absent from the left, such as $\pi(\{B\cup B'\})$; with nonnegative atoms neither makes it smaller. The bounds follow by inserting Eq.~\eqref{eq:coarse} and dropping the overlap that does not appear in each bound.
\end{proof}

Condition (M) says that redundancy with a set of partners cannot shrink when the set grows; Hi-Fi's $\pos{C_{A,\cdot}}$ can, since $C_{A,B\cup B'}=C_{A,B}+C(Y;X_A;X_{B'}\mid X_B)$ drops whenever the added sources are synergistic with $X_A$ given $X_B$. Condition (S) says that information $X_A$ shares with the group $(X_B,X_{B'})$ but with neither part alone must be synergy of $X_B$ and $X_{B'}$. With $\delta=0$ on every pair, both become conditions on the co-informations alone, namely that the right-hand sides of Eqs.~\eqref{eq:dM} and~\eqref{eq:dS} are nonpositive; we say that a distribution \emph{violates} (M) or (S) when one of them is positive. The next two examples violate (M) and (S), respectively.

\begin{example}[A copy and a partner]
\label{ex:copy}
Let $a$, $b$ and $c$ be independent uniform bits, $X_1=(a,b)$, $X_2=a$, $X_3=c$ and $Y=(a,b\oplus c)$. Then $I_3=0$, $I_1=I_2=I_{12}=I_{23}=1$ and $I_{13}=I_{123}=2$ bits, so $C_{1,2}=1$ bit, $C_{1,3}=-1$ bit and $C_{1,23}=0$. Thus, against the merged pair, redundancy with the copy and synergy with the partner cancel, and zero overlap would leave that pair with neither. Condition (M) with $A=\{1\}$, $B=\{2\}$, $B'=\{3\}$ rules this out because every nonnegative PID has $\Red_{1,23}\ge\Red_{1,2}\ge1$ bit and hence $\Syn_{1,23}\ge1$ bit, so this pair cannot have zero overlap even on its own. The other five pairs can: the PID with 1 bit on $\{1\}\{2\}$ and 1 bit on $\{13\}$ has zero overlap on all of them and the smallest possible overlap, 1 bit, on $(X_1,(X_2,X_3))$. The same mechanism appears in linear Gaussian models when $X_2$ is a noisy proxy of $X_1$ and $X_3$ an independent second cause of $Y$.
\end{example}

\begin{example}[The distribution of Rauh \emph{et al.}~\cite{Rauh2014}]
\label{ex:rauh}
Let $X_1$ and $X_2$ be independent uniform bits, $X_3=X_1\oplus X_2$ and $Y=(X_1,X_2,X_3)$, the distribution with which Rauh \emph{et al.}\ showed that the Williams--Beer axioms, local positivity and the identity property cannot hold together. Any two sources determine the third, so $C_{i,j}=0$ and $C_{i,jk}=1$ bit: each source shares its bit with the other two together but with neither alone. Condition (S) rules out zero overlap: the bit that $X_i$ shares with $(X_j,X_k)$ must appear as redundancy of $X_i$ with $X_j$ or with $X_k$, or as synergy of $X_j$ and $X_k$, and all three pairs have zero co-information, so by Eq.~\eqref{eq:dS}, $\delta_{1,2}+\delta_{1,3}+\delta_{2,3}\ge1$ bit. Each overlap can be zero on its own, but not all three.
\end{example}

For three sources the converse of Proposition~\ref{prop:acrosspairs} also holds, so the mechanisms of these two examples are the only ones.

\begin{theorem}[Three sources]
\label{thm:three}
Let $n=3$. For every nondecreasing set function $f$ with $f(\emptyset)=0$, in particular for $f(A)=I(Y;X_A)$ of every distribution, write $C_{A,B}=f(A)+f(B)-f(A\cup B)$, and call a nonnegative solution $\pi$ of Eq.~\eqref{eq:shannon_n} with $f(A)$ in place of $I(Y;X_A)$ a nonnegative PID of $f$ with zero overlap on the pair $(A,B)$ if $I_\cap(\{A,B\})=\pos{C_{A,B}}$. Then a nonnegative PID of $f$ with zero overlap on all six coarse-grained pairs exists if and only if the right-hand sides of Eqs.~\eqref{eq:dM} and~\eqref{eq:dS} are nonpositive for $A=\{i\}$, $B=\{j\}$ and $B'=\{k\}$, for every ordering $(i,j,k)$ of $(1,2,3)$.
\end{theorem}

\begin{proof}
Necessity is Proposition~\ref{prop:acrosspairs}. Sufficiency is proved in Appendix~\ref{app:three}: zero overlap fixes $I_\cap$ at 13 of the 18 nodes, and the five free values can be eliminated by hand. A computation of the feasible cones for all 64 sign patterns of the co-informations with cddlib~\cite{cddlib} agrees.
\end{proof}

Example~\ref{ex:rauh} has an identity target, $Y=(X_1,X_2,X_3)$. For identity targets, whether zero overlap on the pairs that Hi-Fi examines is attainable can be decided for any number of sources.

\begin{proposition}[Identity targets]
\label{prop:identity}
Let $Y=(X_1,\dots,X_n)$. Then $C_{A,B}=I(X_A;X_B)\ge0$ for every pair, so Hi-Fi reports no synergy. A nonnegative PID with zero overlap on every pair that Hi-Fi examines exists if and only if the information diagram $\mu$ of the sources~\cite{Yeung1991} is nonnegative. It is then unique and has zero overlap on every coarse-grained pair, not only on those that Hi-Fi examines: the node $\{\{t\}:t\in T\}$ carries $\mu(T)$, the co-information of $X_T$ given the other sources, and all other atoms are zero. For $n=3$ the condition is $C(X_1;X_2;X_3)\ge0$, and the atoms are $\pi(\{i\})=H(X_i\mid X_j,X_k)$, $\pi(\{i\}\{j\})=I(X_i;X_j\mid X_k)$ and $\pi(\{1\}\{2\}\{3\})=C(X_1;X_2;X_3)$.
\end{proposition}

The proof is in Appendix~\ref{app:prop9}. The natural multivariate counterpart of Hi-Fi is the nonnegative PID that minimizes the total overlap $\sum\delta_{i,Z}$ over the $q(n)$ distinct pair nodes that Hi-Fi examines (each pair of single sources counted once), a linear program in the $D(n)-2$ atoms whose optimum is zero exactly when zero overlap is feasible. We call its minimizers minimal-overlap PIDs and its optimal value the forced overlap. It reduces to Hi-Fi for $n=2$ and, by Proposition~\ref{prop:identity}, to the information diagram for identity targets with a nonnegative diagram, but it is not unique in general, e.g. in Example~\ref{ex:rauh} the forced bit can sit on any one of the three pairs or be spread as $1/3$ bit on each.

For two sources the information diagram is always nonnegative and Proposition~\ref{prop:identity} reduces to the identity property of Hi-Fi, while for three or more it can be negative, and the distribution of Example~\ref{ex:rauh} has $C(X_1;X_2;X_3)=-1$ bit. Consistently with Theorem~\ref{thm:three}, condition (S) there reads $I(X_i;X_j,X_k)\le I(X_i;X_j)+I(X_i;X_k)$, which is $C(X_1;X_2;X_3)\ge0$. Matthias \emph{et al.}~\cite{Matthias2025} have shown that the obstruction of Rauh \emph{et al.}\ survives under weaker assumptions than local positivity, and Lyu \emph{et al.}~\cite{Lyu2025,Lyu2025b,Lyu2026} that for four or more sources no decomposition indexed by the antichain lattice is consistent across all subsets of sources. Proposition~\ref{prop:identity} gives, for identity targets and the overlap formulation, the exact boundary at which the obstruction sets in.

\begin{table*}[t]
\caption{Zero overlap on all coarse-grained pairs that Hi-Fi examines, in random distributions with $n$ sources (generating models and seed in Appendix~\ref{app:numerics}). Infeasible: no nonnegative PID has $\delta_{i,Z}=0$ for all $(i,Z)$. (M), (S) only: share of the infeasible cases that violate (M), and that satisfy (M) but violate (S), evaluated for single-source $A$ and disjoint $B,B'$ among the remaining sources; no infeasible case satisfied both (for the pairs that Hi-Fi examines; the identity rows use Dirichlet concentration 1, see the text for concentration 0.3). Forced overlap: median minimal total overlap over the infeasible cases, relative to $I(Y;X_1,\dots,X_n)$. Reported pairs: zero overlap required only on the pairs $(i,Z_R)$ and $(i,Z_S)$ that exhaustive Hi-Fi reports. $R,S>0$: drivers with both $R_i$ and $S_i$ above $\max(0.05\,I_i,10^{-3})$ bits. All entries in percent; binomial standard errors are at most 1.6 points for $N=1000$ and 3.5 points for $N=200$.}
\label{tab:multi}
\begin{ruledtabular}
\begin{tabular}{llrrrrrrr}
 & Family & $N$ & Infeasible & (M) & (S) only & Forced overlap & Reported pairs & $R,S>0$\\
\hline
\input{tables/tab_multi}
\end{tabular}
\end{ruledtabular}
\end{table*}

Table~\ref{tab:multi} shows that the failure is the rule rather than the exception. Zero overlap on all pairs that Hi-Fi examines was infeasible in 21--79\% of the random three-source distributions and in 51--100\% of the four-source ones. A weaker requirement fails too, namely that in separate samples of 300 random binary and 300 random Gaussian three-source distributions (ensemble G), the smallest overlaps that the pairs can have individually were not attainable by one nonnegative PID in 13\% and 23\% of the cases. Condition (M) accounted for 95--100\% of the failures except for identity targets, where (M) is never violated and (S) accounts for all of them. Even when zero overlap is required only on the two pairs that Hi-Fi reports for each driver, it was infeasible in 10--27\% of the non-identity three-source and 42--68\% of the non-identity four-source distributions. The forced overlap is modest in the median, 2--12\% of $I(Y;X_1,\dots,X_n)$ for three sources, but reached a third of it for four-source Gaussian systems. Additive models are not exempt from this. With correlated binary sources and an additive logistic target, 70\% of the distributions were infeasible, almost all through (M), i.e.\ because the conditional information of some driver first falls and then rises as partners are added.

%% file: tables/tab_count.tex
2 & 4 & 3 & 1 & 1 & 1 & 0 \\
3 & 18 & 7 & 11 & 6 & 6 & 5 \\
4 & 166 & 15 & 151 & 25 & 22 & 129 \\
5 & 7\,579 & 31 & 7\,548 & 90 & 65 & 7\,483 \\
6 & 7\,828\,352 & 63 & 7\,828\,289 & 301 & 171 & 7\,828\,118 \\

%% file: figs/fig_lattice3.tex
\begin{tikzpicture}[font=\scriptsize,
  nd/.style={draw, rounded corners=2pt, inner sep=1.2pt, minimum height=9pt, text=black},
  ed/.style={draw=black!38, line width=0.35pt},
  bold/.style={line width=1.0pt},
  sh/.style={fill=black!11, draw=black!60},
  pr/.style={fill=cB!28, draw=cB},
  un/.style={fill=white, draw=black!55, dashed},
  red/.style={fill=cB!28, draw=cB},
  unA/.style={fill=cA!30, draw=cA!85!black},
  unB/.style={fill=cV!22, draw=cV},
  syn/.style={fill=cO!30, draw=cO},
  drop/.style={fill=white, draw=black!40, dashed, text=black!45},
  sw/.style={draw, rounded corners=1pt, minimum width=8pt, minimum height=6pt, inner sep=0pt},
  ttl/.style={anchor=south, font=\footnotesize}]
  \definecolor{cB}{HTML}{2A78D6}\definecolor{cO}{HTML}{EB6834}
  \definecolor{cA}{HTML}{1BAF7A}\definecolor{cV}{HTML}{4A3AA7}
  \node[nd,un] (an1x2x3x) at (0.00,0.00) {$\{1\}\{2\}\{3\}$};
  \node[nd,pr] (an1x2x) at (-1.42,0.64) {$\{1\}\{2\}$};
  \node[nd,pr] (an1x3x) at (0.00,0.64) {$\{1\}\{3\}$};
  \node[nd,pr] (an2x3x) at (1.42,0.64) {$\{2\}\{3\}$};
  \node[nd,pr] (an1x23x) at (-1.42,1.28) {$\{1\}\{23\}$};
  \node[nd,pr] (an2x13x) at (0.00,1.28) {$\{2\}\{13\}$};
  \node[nd,pr] (an3x12x) at (1.42,1.28) {$\{3\}\{12\}$};
  \node[nd,sh] (an1x) at (-2.05,1.92) {$\{1\}$};
  \node[nd,sh] (an2x) at (-0.65,1.92) {$\{2\}$};
  \node[nd,un] (an12x13x23x) at (0.65,1.92) {$\{12\}\{13\}\{23\}$};
  \node[nd,sh] (an3x) at (2.05,1.92) {$\{3\}$};
  \node[nd,un] (an12x13x) at (-1.42,2.56) {$\{12\}\{13\}$};
  \node[nd,un] (an12x23x) at (0.00,2.56) {$\{12\}\{23\}$};
  \node[nd,un] (an13x23x) at (1.42,2.56) {$\{13\}\{23\}$};
  \node[nd,sh] (an12x) at (-1.42,3.20) {$\{12\}$};
  \node[nd,sh] (an13x) at (0.00,3.20) {$\{13\}$};
  \node[nd,sh] (an23x) at (1.42,3.20) {$\{23\}$};
  \node[nd,sh] (an123x) at (0.00,3.84) {$\{123\}$};
  \draw[ed] (an1x) -- (an12x13x);
  \draw[ed] (an2x) -- (an12x23x);
  \draw[ed] (an3x) -- (an13x23x);
  \draw[ed] (an12x) -- (an123x);
  \draw[ed] (an13x) -- (an123x);
  \draw[ed] (an23x) -- (an123x);
  \draw[ed] (an1x2x) -- (an1x23x);
  \draw[ed] (an1x2x) -- (an2x13x);
  \draw[ed] (an1x3x) -- (an1x23x);
  \draw[ed] (an1x3x) -- (an3x12x);
  \draw[ed] (an1x23x) -- (an1x);
  \draw[ed] (an1x23x) -- (an12x13x23x);
  \draw[ed] (an2x3x) -- (an2x13x);
  \draw[ed] (an2x3x) -- (an3x12x);
  \draw[ed] (an2x13x) -- (an2x);
  \draw[ed] (an2x13x) -- (an12x13x23x);
  \draw[ed] (an3x12x) -- (an3x);
  \draw[ed] (an3x12x) -- (an12x13x23x);
  \draw[ed] (an12x13x) -- (an12x);
  \draw[ed] (an12x13x) -- (an13x);
  \draw[ed] (an12x23x) -- (an12x);
  \draw[ed] (an12x23x) -- (an23x);
  \draw[ed] (an13x23x) -- (an13x);
  \draw[ed] (an13x23x) -- (an23x);
  \draw[ed] (an1x2x3x) -- (an1x2x);
  \draw[ed] (an1x2x3x) -- (an1x3x);
  \draw[ed] (an1x2x3x) -- (an2x3x);
  \draw[ed] (an12x13x23x) -- (an12x13x);
  \draw[ed] (an12x13x23x) -- (an12x23x);
  \draw[ed] (an12x13x23x) -- (an13x23x);
  \node[nd,red] (bn1x2x3x) at (5.95,0.00) {$\{1\}\{2\}\{3\}$};
  \node[nd,red,bold] (bn1x2x) at (4.53,0.64) {$\{1\}\{2\}$};
  \node[nd,unA] (bn1x3x) at (5.95,0.64) {$\{1\}\{3\}$};
  \node[nd,unB] (bn2x3x) at (7.37,0.64) {$\{2\}\{3\}$};
  \node[nd,unA] (bn1x23x) at (4.53,1.28) {$\{1\}\{23\}$};
  \node[nd,unB] (bn2x13x) at (5.95,1.28) {$\{2\}\{13\}$};
  \node[nd,syn] (bn3x12x) at (7.37,1.28) {$\{3\}\{12\}$};
  \node[nd,unA,bold] (bn1x) at (3.90,1.92) {$\{1\}$};
  \node[nd,unB,bold] (bn2x) at (5.30,1.92) {$\{2\}$};
  \node[nd,syn] (bn12x13x23x) at (6.60,1.92) {$\{12\}\{13\}\{23\}$};
  \node[nd,drop] (bn3x) at (8.00,1.92) {$\{3\}$};
  \node[nd,syn] (bn12x13x) at (4.53,2.56) {$\{12\}\{13\}$};
  \node[nd,syn] (bn12x23x) at (5.95,2.56) {$\{12\}\{23\}$};
  \node[nd,drop] (bn13x23x) at (7.37,2.56) {$\{13\}\{23\}$};
  \node[nd,syn,bold] (bn12x) at (4.53,3.20) {$\{12\}$};
  \node[nd,drop] (bn13x) at (5.95,3.20) {$\{13\}$};
  \node[nd,drop] (bn23x) at (7.37,3.20) {$\{23\}$};
  \node[nd,drop] (bn123x) at (5.95,3.84) {$\{123\}$};
  \draw[ed] (bn1x) -- (bn12x13x);
  \draw[ed] (bn2x) -- (bn12x23x);
  \draw[ed] (bn3x) -- (bn13x23x);
  \draw[ed] (bn12x) -- (bn123x);
  \draw[ed] (bn13x) -- (bn123x);
  \draw[ed] (bn23x) -- (bn123x);
  \draw[ed] (bn1x2x) -- (bn1x23x);
  \draw[ed] (bn1x2x) -- (bn2x13x);
  \draw[ed] (bn1x3x) -- (bn1x23x);
  \draw[ed] (bn1x3x) -- (bn3x12x);
  \draw[ed] (bn1x23x) -- (bn1x);
  \draw[ed] (bn1x23x) -- (bn12x13x23x);
  \draw[ed] (bn2x3x) -- (bn2x13x);
  \draw[ed] (bn2x3x) -- (bn3x12x);
  \draw[ed] (bn2x13x) -- (bn2x);
  \draw[ed] (bn2x13x) -- (bn12x13x23x);
  \draw[ed] (bn3x12x) -- (bn3x);
  \draw[ed] (bn3x12x) -- (bn12x13x23x);
  \draw[ed] (bn12x13x) -- (bn12x);
  \draw[ed] (bn12x13x) -- (bn13x);
  \draw[ed] (bn12x23x) -- (bn12x);
  \draw[ed] (bn12x23x) -- (bn23x);
  \draw[ed] (bn13x23x) -- (bn13x);
  \draw[ed] (bn13x23x) -- (bn23x);
  \draw[ed] (bn1x2x3x) -- (bn1x2x);
  \draw[ed] (bn1x2x3x) -- (bn1x3x);
  \draw[ed] (bn1x2x3x) -- (bn2x3x);
  \draw[ed] (bn12x13x23x) -- (bn12x13x);
  \draw[ed] (bn12x13x23x) -- (bn12x23x);
  \draw[ed] (bn12x13x23x) -- (bn13x23x);
  \node[nd,red] (cn1x2x3x) at (11.90,0.00) {$\{1\}\{2\}\{3\}$};
  \node[nd,red] (cn1x2x) at (10.48,0.64) {$\{1\}\{2\}$};
  \node[nd,red] (cn1x3x) at (11.90,0.64) {$\{1\}\{3\}$};
  \node[nd,unB] (cn2x3x) at (13.32,0.64) {$\{2\}\{3\}$};
  \node[nd,red,bold] (cn1x23x) at (10.48,1.28) {$\{1\}\{23\}$};
  \node[nd,unB] (cn2x13x) at (11.90,1.28) {$\{2\}\{13\}$};
  \node[nd,unB] (cn3x12x) at (13.32,1.28) {$\{3\}\{12\}$};
  \node[nd,unA,bold] (cn1x) at (9.85,1.92) {$\{1\}$};
  \node[nd,unB] (cn2x) at (11.25,1.92) {$\{2\}$};
  \node[nd,unB] (cn12x13x23x) at (12.55,1.92) {$\{12\}\{13\}\{23\}$};
  \node[nd,unB] (cn3x) at (13.95,1.92) {$\{3\}$};
  \node[nd,syn] (cn12x13x) at (10.48,2.56) {$\{12\}\{13\}$};
  \node[nd,unB] (cn12x23x) at (11.90,2.56) {$\{12\}\{23\}$};
  \node[nd,unB] (cn13x23x) at (13.32,2.56) {$\{13\}\{23\}$};
  \node[nd,syn] (cn12x) at (10.48,3.20) {$\{12\}$};
  \node[nd,syn] (cn13x) at (11.90,3.20) {$\{13\}$};
  \node[nd,unB,bold] (cn23x) at (13.32,3.20) {$\{23\}$};
  \node[nd,syn,bold] (cn123x) at (11.90,3.84) {$\{123\}$};
  \draw[ed] (cn1x) -- (cn12x13x);
  \draw[ed] (cn2x) -- (cn12x23x);
  \draw[ed] (cn3x) -- (cn13x23x);
  \draw[ed] (cn12x) -- (cn123x);
  \draw[ed] (cn13x) -- (cn123x);
  \draw[ed] (cn23x) -- (cn123x);
  \draw[ed] (cn1x2x) -- (cn1x23x);
  \draw[ed] (cn1x2x) -- (cn2x13x);
  \draw[ed] (cn1x3x) -- (cn1x23x);
  \draw[ed] (cn1x3x) -- (cn3x12x);
  \draw[ed] (cn1x23x) -- (cn1x);
  \draw[ed] (cn1x23x) -- (cn12x13x23x);
  \draw[ed] (cn2x3x) -- (cn2x13x);
  \draw[ed] (cn2x3x) -- (cn3x12x);
  \draw[ed] (cn2x13x) -- (cn2x);
  \draw[ed] (cn2x13x) -- (cn12x13x23x);
  \draw[ed] (cn3x12x) -- (cn3x);
  \draw[ed] (cn3x12x) -- (cn12x13x23x);
  \draw[ed] (cn12x13x) -- (cn12x);
  \draw[ed] (cn12x13x) -- (cn13x);
  \draw[ed] (cn12x23x) -- (cn12x);
  \draw[ed] (cn12x23x) -- (cn23x);
  \draw[ed] (cn13x23x) -- (cn13x);
  \draw[ed] (cn13x23x) -- (cn23x);
  \draw[ed] (cn1x2x3x) -- (cn1x2x);
  \draw[ed] (cn1x2x3x) -- (cn1x3x);
  \draw[ed] (cn1x2x3x) -- (cn2x3x);
  \draw[ed] (cn12x13x23x) -- (cn12x13x);
  \draw[ed] (cn12x13x23x) -- (cn12x23x);
  \draw[ed] (cn12x13x23x) -- (cn13x23x);
  \node[ttl] at (0.00,4.14) {\textbf{(a)} the redundancy lattice for $n=3$};
  \node[ttl] at (5.95,4.14) {\textbf{(b)} the pair $A=\{1\}$, $B=\{2\}$};
  \node[ttl] at (11.90,4.14) {\textbf{(c)} the pair $A=\{1\}$, $B=\{23\}$};
  \node[sw,sh] at (-2.35,-0.55) {};
  \node[anchor=west] at (-2.18,-0.55) {$\{A\}$: $I_\cap=I(Y;X_A)$, Eq.~\eqref{eq:shannon_n} (7 nodes)};
  \node[sw,pr] at (-2.35,-0.91) {};
  \node[anchor=west] at (-2.18,-0.91) {$\{A\}\{B\}$, $A\cap B=\emptyset$: pair, $I_\cap=\Red_{A,B}$ (6)};
  \node[sw,un] at (-2.35,-1.27) {};
  \node[anchor=west] at (-2.18,-1.27) {in no pair (5)};
  \node[sw,red] at (4.35,-0.55) {};
  \node[anchor=west] at (4.52,-0.55) {$\Red_{1,2}$};
  \node[sw,unA] at (6.25,-0.55) {};
  \node[anchor=west] at (6.42,-0.55) {$\Un_{1}$};
  \node[sw,unB] at (4.35,-0.91) {};
  \node[anchor=west] at (4.52,-0.91) {$\Un_{2}$};
  \node[sw,syn] at (6.25,-0.91) {};
  \node[anchor=west] at (6.42,-0.91) {$\Syn_{1,2}$};
  \node[sw,drop] at (4.35,-1.27) {};
  \node[anchor=west] at (4.52,-1.27) {dropped ($A\cup B\notin\mathcal U(\alpha)$)};
  \node[sw,red] at (10.30,-0.55) {};
  \node[anchor=west] at (10.47,-0.55) {$\Red_{1,23}$};
  \node[sw,unA] at (12.20,-0.55) {};
  \node[anchor=west] at (12.37,-0.55) {$\Un_{1}$};
  \node[sw,unB] at (10.30,-0.91) {};
  \node[anchor=west] at (10.47,-0.91) {$\Un_{23}$};
  \node[sw,syn] at (12.20,-0.91) {};
  \node[anchor=west] at (12.37,-0.91) {$\Syn_{1,23}$};
  \node[anchor=west, font=\footnotesize] at (-2.45,-1.95) {\textbf{(d)} Hi-Fi's pairs: 9 combinations $(i,Z)$, 6 distinct nodes};
  \node[anchor=west, align=left, text=black!70] at (-2.45,-2.75) {driver $i$,\\partner set $Z$};
  \node[anchor=west, align=left, text=black!70] at (-2.45,-3.65) {node $\{i,Z\}$};
  \node[text=black!70] at (2.00,-2.33) {$i=1$};
  \draw[black!45, line width=0.4pt] (0.20,-2.47) -- (3.80,-2.47);
  \node[inner sep=1.2pt] (h1x2) at (0.55,-2.75) {$(1,\{2\})$};
  \node[inner sep=1.2pt] (h1x3) at (2.00,-2.75) {$(1,\{3\})$};
  \node[inner sep=1.2pt] (h1x23) at (3.45,-2.75) {$(1,\{23\})$};
  \node[text=black!70] at (6.70,-2.33) {$i=2$};
  \draw[black!45, line width=0.4pt] (4.90,-2.47) -- (8.50,-2.47);
  \node[inner sep=1.2pt] (h2x1) at (5.25,-2.75) {$(2,\{1\})$};
  \node[inner sep=1.2pt] (h2x3) at (6.70,-2.75) {$(2,\{3\})$};
  \node[inner sep=1.2pt] (h2x13) at (8.15,-2.75) {$(2,\{13\})$};
  \node[text=black!70] at (11.40,-2.33) {$i=3$};
  \draw[black!45, line width=0.4pt] (9.60,-2.47) -- (13.20,-2.47);
  \node[inner sep=1.2pt] (h3x1) at (9.95,-2.75) {$(3,\{1\})$};
  \node[inner sep=1.2pt] (h3x2) at (11.40,-2.75) {$(3,\{2\})$};
  \node[inner sep=1.2pt] (h3x12) at (12.85,-2.75) {$(3,\{12\})$};
  \node[nd,pr] (hn1x2x) at (2.25,-3.65) {$\{1\}\{2\}$};
  \node[nd,pr] (hn1x23x) at (3.95,-3.65) {$\{1\}\{23\}$};
  \node[nd,pr] (hn1x3x) at (5.90,-3.65) {$\{1\}\{3\}$};
  \node[nd,pr] (hn2x13x) at (7.85,-3.65) {$\{2\}\{13\}$};
  \node[nd,pr] (hn2x3x) at (9.80,-3.65) {$\{2\}\{3\}$};
  \node[nd,pr] (hn3x12x) at (12.85,-3.65) {$\{3\}\{12\}$};
  \draw[ed, black!55, -{Stealth[length=3pt]}] (h1x2.south) -- (hn1x2x.north);
  \draw[ed, black!55, -{Stealth[length=3pt]}] (h1x3.south) -- (hn1x3x.north);
  \draw[ed, black!55, -{Stealth[length=3pt]}] (h1x23.south) -- (hn1x23x.north);
  \draw[ed, black!55, -{Stealth[length=3pt]}] (h2x1.south) -- (hn1x2x.north);
  \draw[ed, black!55, -{Stealth[length=3pt]}] (h2x3.south) -- (hn2x3x.north);
  \draw[ed, black!55, -{Stealth[length=3pt]}] (h2x13.south) -- (hn2x13x.north);
  \draw[ed, black!55, -{Stealth[length=3pt]}] (h3x1.south) -- (hn1x3x.north);
  \draw[ed, black!55, -{Stealth[length=3pt]}] (h3x2.south) -- (hn2x3x.north);
  \draw[ed, black!55, -{Stealth[length=3pt]}] (h3x12.south) -- (hn3x12x.north);
\end{tikzpicture}

%% file: tables/tab_multi.tex
$n=3$ & binary target, Dirichlet(1) & 1000 & 41 & 100 & 0 & 7 & 23 & 22 \\
 & binary target, Dirichlet(0.2) & 1000 & 61 & 95 & 5 & 9 & 14 & 35 \\
 & ternary target, Dirichlet(1) & 1000 & 21 & 100 & 0 & 6 & 12 & 11 \\
 & logistic, additive & 1000 & 70 & 95 & 5 & 4 & 10 & 22 \\
 & logistic, with interactions & 1000 & 70 & 100 & 0 & 12 & 23 & 28 \\
 & Gaussian, Wishart df $=6$ & 1000 & 76 & 96 & 4 & 7 & 27 & 41 \\
 & Gaussian, Wishart df $=24$ & 1000 & 78 & 97 & 3 & 3 & 16 & 20 \\
 & identity target $Y=\mathbf X$ & 1000 & 79 & 0 & 100 & 2 & 0 & 0 \\
\hline
$n=4$ & binary target, Dirichlet(1) & 200 & 51 & 100 & 0 & 11 & 42 & 22 \\
 & Gaussian, Wishart df $=7$ & 200 & 98 & 100 & 0 & 33 & 68 & 67 \\
 & identity target $Y=\mathbf X$ & 200 & 100 & 0 & 100 & 14 & 0 & 0 \\

%% file: sec8_discussion.tex
\section{Discussion}
\label{sec:discussion}

\subsection{What a behavioral decomposition can claim}

We can now answer the question raised in the Introduction, what these numbers measure. PID, Hi-Fi and the O-information~\cite{Rosas2019} are statements about a joint distribution, about behavior rather than mechanism~\cite{Rosas2022NatPhys}, and the results above sort the claims they support into four levels, each requiring strictly more than the one below. Level 0, which is fully identified, requires the least. It contains the net balance $C$, Hi-Fi's $R$ and $S$ and the bounds of Corollary~\ref{cor:bounds} which are functions of the three Shannon terms alone (of Shannon invariants in the sense of Ref.~\cite{Gutknecht2025}). Level 1 is measure-relative and contains redundancy and synergy values and atom fractions, which, whenever $\dmax>0$, rest on the convention or the reference family of the chosen measure. Every reference-family overlap is of the same kind, the excess of $P$ over the most redundant member of its family (Definition~\ref{def:family}): BROJA's overlap is a statement about $P$ relative to $\DeltaP$, $\Idep$'s relative to $\Ftwo\cup\{\QCI\}$, MMI's relative to a family that keeps only two numbers. Level 2 requires testing model classes. For example, $\Ithree>0$, that no pairwise log-linear model fits, is also a statement about $P$ relative to a family, but $\Ftwo$ is an exponential family and $\Ithree$ a divergence to it, so the claim comes with the standard likelihood-ratio test, and this, not the family form as such, is what separates level 2 from level 1. Level 3 describes mechanism and a claim that two sources act jointly on the target require assumptions outside the distribution, namely no hidden common causes or selection, which create three-way terms among observed variables, a scale on which ``no interaction'' is defined, and a causal order. Hi-Fi and MMI use only level-0 information and differ only in the convention they apply at level 1.

\subsection{Is zero overlap the right default?}

Not in general, because the overlap is not identified by the Shannon terms, and $\delta=0$ is a matter of convention. Hi-Fi fails on the triadic distribution of Sec.~\ref{sec:pairref}, where real redundancy and real synergy coexist, and MMI on the dyadic one. Which end of the segment is conservative depends on the claim (Sec.~\ref{sec:overlap}): $\delta=0$ for claims of redundancy or synergy, $\delta=\dmax$ for claims of unique information, the claim a feature-importance method makes. Synergy above $\pos{-C}$ rests on whatever supports the overlap. For MMI that is the min rule alone, except where MMI coincides with BROJA, as on Gaussian triplets, and MMI produces synergy in 86\% of the additive binary mechanisms of Table~\ref{tab:mmi}. The overlaps of BROJA and $\Idep$ rest on their reference families and are measure-relative in the sense of level 1. For claims of synergy, then, $\delta=0$ is the conservative default, and the pairwise-reference PID refines it by departing from zero only when the distribution shows pairwise redundancy and triadic synergy at the same time, and then by the smaller of the two.

\subsection{Interaction depends on the scale}

Whether two sources interact depends on the scale on which ``no interaction'' is defined. For a strictly positive distribution with a binary target, $\Ithree=0$ exactly when the logit of $P(Y=1\mid x_1,x_2)$ is additive, so triadic synergy is interaction on the logit scale. With uniform inputs the deterministic AND gate has $\Ithree=0$, because its bivariate marginals determine the joint distribution. It is the limit $\beta\to\infty$ of the threshold model $Y\sim\mathrm{Bernoulli}\,\sigma\big(\beta(x_1+x_2-1.5)\big)$, whose logit is additive, so that $\Ithree=0$ for every $\beta$ (Proposition~\ref{prop:pairref}(ii); numerically below $10^{-4}$ bits up to $\beta=40$). Along this path MMI's synergy grows from 0.04 bits at $\beta=1$ to 0.50 bits and $\pos{-C}$ from 0.002 to 0.19 bits, so neither measures interaction on this scale. The noise model matters as well because in the threshold model the noise acts on the additive score, whereas if the output of the deterministic gate is flipped with probability $\epsilon$, the logit is no longer additive and $\Ithree=0.057$, 0.077, 0.066, 0.025 and 0.004 bits at $\epsilon=0.2$, 0.1, 0.05, 0.01 and 0.001, vanishing in the deterministic limit. On the probability scale the answer changes again. The structural synergy of Ref.~\cite{Antonacci2026}, the part of the variance of $E[Y\mid X_1,X_2]$ that no additive predictor $h_1(X_1)+h_2(X_2)$ captures, is $1/16$ for the AND gate and in general positive for the additive logistic family of Sec.~\ref{sec:mmi}, although both have $\Ithree=0$. This is the removable-interaction problem of regression~\cite{Loftus1978}, seen through information theory.

\subsection{Recommendations}

In summary, Hi-Fi's $R$ and $S$ are lower bounds on redundancy and synergy and its $U_i$ are upper bounds on unique information under every nonnegative PID (signed measures excluded). The pairwise-reference PID adds exactly the triadic part, while everything above that is a choice of measure. To support clarity and interpretation, we suggest some good practices when applying Hi-Fi or any nonnegative PID. For triplets, reporting $C$, $\Itwo$, $\Ithree$ with its $G^2$ $p$-value, and $\dmax$ is sufficient. These give Hi-Fi's atoms, the pairwise-reference atoms, and the range of every nonnegative PID. It is important to match the language to the evidence, e.g.\ ``net synergy'' for $\pos{-C}$ (level 0), ``triadic synergy'' for a significant $\Ithree$ (level 2), a mechanism only with design assumptions (level 3). MMI synergy fractions support none of these on their own, and BROJA's support a level-1 claim relative to $\DeltaP$, and when either is reported, $\dmax$ should be reported with it.

For Gaussian or continuous Gaussian-copula data, the common case in neuroscience and physiology, the whole analysis reduces to the sign of $C$ and the length $\dmax$, since $\Ithree=0$, the pairwise-reference PID equals Hi-Fi, and BROJA, $\Imin$, $\Ired$ and MMI coincide at $\dmax$ (Proposition~\ref{prop:gauss}). Discretized Gaussian-copula data, such as Likert items, are not covered by Proposition~\ref{prop:gauss}(i), since discretization is not invertible and the discretized distribution can carry a three-way term, so $\Ithree$ must then be estimated and tested like any other discrete quantity.

With more than two sources, for each driver the partner sets $Z_R$ and $Z_S$ with $R_i$ and $S_i$ could be reported, reading $R_i$ and $S_i$ as lower bounds on coarse-grained redundancy and synergy and $U_i$ as an upper bound on unique information (Theorem~\ref{thm:extremes}). Note that $R_i$ and $S_i$ of one driver usually refer to different partner sets and are not parts of one decomposition, so Hi-Fi atoms should not be summed across drivers or partner sets. Also note that the minimal-overlap linear program of Sec.~\ref{sec:multi} is a diagnostic, not a recommended decomposition, because its minimizer is not unique. When using plug-in estimates, explicit bias control is important: the bias of the co-information of three $L$-level variables is $-(L-1)^3/(2N\ln2)$ bits (Sec.~\ref{sec:pairref}), which can exceed the co-information of weakly correlated variables.

\subsection{Limitations and outlook}

The exact results of Secs.~\ref{sec:two}--\ref{sec:mmi} concern two sources; Sec.~\ref{sec:multi} shows that they hold for every coarse-grained pair, that MMI and Hi-Fi's unique atoms remain extremal, and that zero overlap generally fails jointly, on identity targets exactly when the sources' information diagram has a negative atom, which is what makes the example of Rauh \emph{et al.}~\cite{Rauh2014} work. A canonical choice among the minimal-overlap PIDs of Sec.~\ref{sec:multi}, a characterization of feasibility for $n\ge4$, and an order-resolved picture based on the connected informations $I^{(k)}$~\cite{Amari2001,Schneidman2003} are the natural next steps. All results are population statements about the information-theoretic version of Hi-Fi with exhaustive search, while the greedy search of applications returns valid but possibly looser bounds and can miss an entire bit (Remark~\ref{rem:greedy}).

%% file: appA.tex
\section{Closed form of the dependency-constraint redundancy}
\label{app:idep}

James, Emenheiser and Crutchfield~\cite{James2019Idep} define the unique information of $X_1$ through the lattice of constraint sets on the marginals of $(X_1,X_2,Y)$. For three variables the nodes are the sets of pairwise-marginal constraints, i.e.\ the subsets of $\{X_1X_2,X_1Y,X_2Y\}$ with the univariate marginals always imposed, together with the full joint; each node is assigned the maximum-entropy distribution under its constraints, and an edge adds one constraint. $\Un_1$ is the smallest increase of $I(Y;X_1,X_2)$ along an edge that adds the constraint $\{X_1Y\}$, and the four subsets that do not contain $X_1Y$ give exactly the four such edges. The resulting reference-family form (Proposition~\ref{prop:named}) is specific to two sources. Kay and Ince~\cite{KayInce2018} wrote this as a minimum over the four such edges and evaluated it for Gaussians; we only rewrite their four terms through co-informations. The edges are
\begin{enumerate}
\item $\{X_1\}\{X_2\}\{Y\}\to\{X_1Y\}\{X_2\}$: the joint information rises from 0 to $I_1$.
\item $\{X_1\}\{X_2Y\}\to\{X_1Y\}\{X_2Y\}$: the maximum-entropy distribution at the upper node is $\QCI$, so the increase is $I_{\QCI}(Y;X_1,X_2)-I_2$.
\item $\{X_1X_2\}\{Y\}\to\{X_1X_2\}\{X_1Y\}$: at the upper node $Y\perp X_2\mid X_1$, so the information rises from 0 to $I_1$.
\item $\{X_1X_2\}\{X_2Y\}\to\{X_1X_2\}\{X_1Y\}\{X_2Y\}$: the upper node is $\Ptwo$, so the increase is $I_{\Ptwo}(Y;X_1,X_2)-I_2$.
\end{enumerate}
Since $Q\in\{\QCI,\Ptwo\}$ keeps $I_Q(Y;X_i)=I_i$, each of the last two terms equals $I_Q(Y;X_1\mid X_2)=I_1-C_Q$, and
\begin{equation}
\begin{aligned}
\Red_{\Idep}=I_1-\Un_1&=\max\{0,\,\Itwo,\,C_{\QCI}\}\\
&=\max\{\Itwo,\,I_{\QCI}(X_1;X_2)\},
\end{aligned}
\end{equation}
using $C_{\QCI}=I_{\QCI}(X_1;X_2)\ge0$. The expression is symmetric in the sources, as it must be. It agreed with \texttt{dit}'s \texttt{PID\_dep} to $3\times10^{-10}$ bits on the gates and on dense random distributions, and to $3.5\times10^{-4}$ bits on sparse ones, where the maximum-entropy optimizer of the package limits the precision. For Gaussian variables it reduces to Proposition~\ref{prop:gauss}(iii). The term $I_{\QCI}(X_1;X_2)$ by itself is the closed-form redundancy of Lyu \emph{et al.}~\cite{Lyu2025} (Sec.~\ref{sec:measures}).

\section{Numerical methods and random ensembles}
\label{app:numerics}

All discrete quantities were computed exactly from probability tables. $\Ptwo$ was obtained by iterative proportional fitting~\cite{DarrochRatcliff1972} to a tolerance of $10^{-13}$ on the three bivariate marginals, and $\Ithree=H(\Ptwo)-H(P)$. BROJA was computed by solving $\max_{Q\in\DeltaP}H_Q(Y\mid X_1,X_2)$ as a convex program with relative-entropy cones in CVXPY~\cite{Diamond2016cvxpy} with the Clarabel interior-point solver~\cite{Goulart2024Clarabel}; the BROJA optimizer of the \texttt{dit} package~\cite{dit2018} stopped more than $10^{-3}$ bits short of the optimum in 24\%, 26\% and 65\% of the three parts of ensemble A, by up to 0.21 bits, and in 10--24\% of them returned a redundancy below $\pos{C}$, which is a negative atom that BROJA cannot produce (see also Ref.~\cite{Makkeh2018BROJA}). With the exact solver, BROJA reported $\Red>0$ and $\Syn>0$ together in 99--100\% of the distributions of ensemble A when both atoms are required to exceed $10^{-6}$ bits, in 91--100\% at $10^{-4}$ bits and in 74--99\% at $10^{-3}$ bits, because on binary tables one of BROJA's atoms is often small. $\Imin$, $\Idep$, MMI, $\Iccs$, $\Ipm$, $\Isx$, $I_{\mathrm{IG}}$, $I_{\mathrm{RAV}}$, $I_{\mathrm{RA}}$, $\Irr$ and $\IGK$ were taken from \texttt{dit} 2.3. Gaussian information quantities follow from log-determinants of correlation submatrices; $\Imin$ was checked against MMI by quadrature of the specific information (agreement better than $10^{-12}$ bits), and $\Iccs$ was estimated by Monte Carlo with $2\times10^6$ draws using the local co-information with the sign condition of Ref.~\cite{Ince2017ccs}, consistent with \texttt{dit}'s \texttt{PID\_CCS} on discretizations of the same Gaussians.

Table~\ref{tab:ensembles} lists every random ensemble used in the paper, with the seed of the generator (\texttt{numpy} \texttt{default\_rng}). Dirichlet ensembles draw the joint probability table from a symmetric Dirichlet distribution with the stated concentration. The additive-logistic family draws an input law $P(X_1,X_2)\sim\mathrm{Dirichlet}(1)$, $a\sim N(0,1)$ and $b_i\sim N(0,1.5^2)$, and sets $P(Y=1\mid x_1,x_2)=\sigma(a+b_1x_1+b_2x_2)$; the interaction family adds $c\,x_1x_2$ with $c\sim N(0,3^2)$. Gaussian triplets use correlation matrices drawn uniformly from the admissible set (rejection sampling). In Table~\ref{tab:mmi}, triplets with $I_{12}\le10^{-4}$ bits were excluded before computing synergy fractions, which is why $n$ is 599 and 19\,999 instead of 600 and 20\,000. For Sec.~\ref{sec:multi}, the $n$-source families draw the joint table of $(Y,X_1,\dots,X_n)$ from a Dirichlet distribution (binary or ternary target, binary sources), a correlation matrix from a normalized Wishart distribution with the stated degrees of freedom, an identity target from a Dirichlet law on the $2^n$ source configurations, or binary sources from an Ising law with couplings $N(0,0.8^2)$ and fields $N(0,0.3^2)$ with a logistic target whose coefficients are $N(0,1.5^2)$ (plus pairwise and $n$-way product terms with $N(0,1.5^2)$ coefficients in the interacting version).

\begin{table}[tb]
\caption{Random ensembles. Part: the figure, table or section that uses it. Size per part and Dirichlet concentration (or Wishart degrees of freedom) in parentheses.}
\label{tab:ensembles}
\begin{ruledtabular}
\footnotesize
\begin{tabular}{@{}p{0.08\columnwidth}p{0.3\columnwidth}p{0.4\columnwidth}p{0.16\columnwidth}@{}}
Name & Used in & Composition & Seed\\
\hline
A & Sec.~\ref{sec:numerics} & binary 300 (1) $+$ 300 (0.2); ternary 200 (1) & 20261001\\
B & Sec.~\ref{sec:pairref} & binary 400 (1) $+$ 300 (0.3); ternary 120 (1) & 21\\
C & Table~\ref{tab:mmi}, Sec.~\ref{sec:numerics} & additive logistic 600 and with interaction 600 (seed 11); Gaussian 20\,000 (seed 12) & 11, 12\\
D & $G^2$ test, Sec.~\ref{sec:pairref} & 3 additive models $\times$ 400 replicates, $N=2000$ & 31\\
E & Table~\ref{tab:multi} & as listed there ($N=1000$ or 200 per family) & 20261006\\
F & Prop.~\ref{prop:identity} check & identity targets, 300 each for $n=3,4$ (seed 404), plus the $n=3$ identity row of E & 404\\
G & Sec.~\ref{sec:multi}, individual minima & three sources, binary 300 (concentration 0.2 or 1, drawn at random per sample) $+$ Gaussian 300 (Wishart df 6); same generator stream as F & 404\\
H & App.~\ref{app:further} & Gaussian triplets 2000, $\Iccs$ on the first 400 & 7\\
\end{tabular}
\end{ruledtabular}
\end{table}

\emph{The $G^2$ test.} On three additive binary models drawn at random from the additive-logistic family (ensemble D, 400 replicates each, $N=2000$), the rejection rates at $\alpha=0.05$ were 0.07, 0.06 and 0.035 (Monte Carlo standard error 0.011; the first lies 1.8 standard errors above nominal), and the mean plug-in $\hat I^{(3)}$ matched its expected value $1/(2N\ln2)$. Power was 1.00 for interaction models with logit $-1+0.5x_1+0.5x_2+c\,x_1x_2$, uniform inputs and $c\in\{1,2,3\}$, for which $\Ithree\ge0.0097$ bits.

\emph{Verification.} The central results (Table~\ref{tab:gates}, the ordering of Eq.~\eqref{eq:order}, Eq.~\eqref{eq:delta2}, the Gaussian structures, the direct-edge example of Sec.~\ref{sec:mmi} and the families of Table~\ref{tab:mmi}) were recomputed with an independent implementation written from the definitions, including a separate BROJA solver and maximum-entropy fit; all agreed to the stated precision, and the simulated families, rerun with a different seed, within the standard errors stated in the caption of Table~\ref{tab:mmi}. For Sec.~\ref{sec:multi}, the lattice counts, Lemma~\ref{lem:layercake}, Theorem~\ref{thm:extremes}, Proposition~\ref{prop:identity} and both examples were re-checked by a second implementation, and Theorem~\ref{thm:three} by an exact rational computation over all 64 sign patterns. Feasibility of the minimal-overlap linear program agreed with the criterion of Proposition~\ref{prop:identity} for all of 1300 random three-source and 300 random four-source identity targets (ensembles E and F), and for three sources the minimal-overlap PID matched the information diagram to $7\times10^{-16}$ bits whenever it was feasible.

\section{Further measures on the segment}
\label{app:further}

On Gaussian triplets with correlation matrices drawn uniformly from the admissible set (ensemble H), $\lambda_{\Idep}$ had median 0.18, was exactly 0 in 77\% of the net-redundant triplets and never 0 in the net-synergistic ones, and $\Iccs$ left the segment in 46\% of the 400 draws for which it was computed.

The auxiliary-variable redundancy $I_{\mathrm{RAV}}$ was contributed by J.~Emenheiser to the \texttt{dit} package~\cite{dit2018} and, to our knowledge, has not been published separately. It is not known to be of reference-family form, but a direct construction places it between Hi-Fi and BROJA.

\begin{proposition}
\label{prop:rav}
Let $\Red_{\mathrm{RAV}}=\max_fI(X_1{:}X_2{:}Y{:}f(X_1,X_2))$, where $f$ ranges over deterministic functions of the sources and $I({\cdot}{:}{\cdot}{:}{\cdot}{:}{\cdot})$ is the four-variable co-information. Then $0\le\delta_{\mathrm{RAV}}\le\delta_{\mathrm{BROJA}}$.
\end{proposition}

\begin{proof}
Write $T=f(X_1,X_2)$. Because $T$ is a function of the sources, $I(X_1;Y\mid X_2)=I(T;Y\mid X_2)+I(X_1;Y\mid X_2,T)$, and expanding $I(X_1{:}X_2{:}Y{:}T)=C-I(X_1{:}X_2{:}Y\mid T)$ gives
\begin{equation}
I(X_1{:}X_2{:}Y{:}T)=I_1-I(X_1;Y\mid T)-I(T;Y\mid X_2).
\label{eq:rav}
\end{equation}
For the lower bound, a constant $T$ gives $0$ and $T=(X_1,X_2)$ gives $C$, so $\Red_{\mathrm{RAV}}\ge\pos{C}$. For the upper bound let $Q(x_1,x_2,y,t)=P(t,y)P(x_1\mid t,y)P(x_2\mid t,y)$. It has the $(T,X_i,Y)$ marginals of $P$, so its $(X_1,X_2,Y)$ marginal lies in $\DeltaP$, and under $Q$ the sources are conditionally independent given $(T,Y)$. Hence $I_Q(Y;X_1\mid X_2)\le I_Q(Y;T\mid X_2)+I_Q(Y;X_1\mid X_2,T)\le I(Y;T\mid X_2)+I(Y;X_1\mid T)$, where the second step uses $I_Q(Y;X_1\mid X_2,T)\le I_Q(Y;X_1\mid T)$, which follows from the conditional independence, and the shared marginals. Since $I_Q(Y;X_1)=I_1$, Eq.~\eqref{eq:rav} gives $C_Q=I_1-I_Q(Y;X_1\mid X_2)\ge I(X_1{:}X_2{:}Y{:}T)$. Maximizing over $f$ gives $\Red_{\mathrm{RAV}}\le\Red_{\mathrm{BROJA}}$.
\end{proof}

On AND, $I_{\mathrm{RAV}}$ lies above $\Idep$ ($\lambda=0.39$ versus $0.26$), and on SUM below it ($0$ versus $0.38$). The information-geometric $I_{\mathrm{IG}}$~\cite{NiuQuinn2019} falls on both sides of BROJA: $\lambda=0.23$ against BROJA's $0$ on COPY, corr., and $0.27$ against BROJA's $1$ on AND. The rescaled redundancy $\Irr$ of Goodwell and Kumar~\cite{GoodwellKumar2017} has an explicit position, $\lambda=I(X_1;X_2)/\min\{H(X_1),H(X_2)\}$: it moves from $R_{\min}=\pos{C}$ toward $R_{\mathrm{MMI}}$ as the coupling between the sources grows, from $\lambda=0$ for independent sources to $\lambda=1$ when one source is a function of the other (the closed form agreed with \texttt{dit} to $7\times10^{-16}$ bits). That coupling is not one of the three Shannon terms, so $\Irr$ is not a rule with constant position. Such a rule, $\Red_\lambda=\pos{C}+\lambda\dmax$ with fixed $\lambda\in[0,1]$, defines a nonnegative PID that is symmetric, self-redundant and monotone for every $\lambda$, but has the identity property only for $\lambda=0$: for $Y=(X_1,X_2)$, $\Red=I(X_1;X_2)=C$ requires $\delta=0$, whereas $\dmax=\min\{H(X_1\mid X_2),H(X_2\mid X_1)\}>0$ unless one source determines the other; BROJA and $\Idep$ satisfy the identity property with positions that vary with $P$ (Sec.~\ref{sec:measures}). $\Irr$ is not ordered relative to the reference-family measures either: for two copies of correlated bits, $Y=(X_1,X_2)$ with $P(X_1=X_2)=0.8$, it has $\lambda=0.28$ while BROJA has $0$, and for $Y=(X_1,X_2)$ it violates the identity property whenever the sources are dependent but neither determines the other.

\section{Gaussian and Ising tables; composition of MMI's synergy}
\label{app:tables}

Table~\ref{tab:gauss} lists the Gaussian structures used in the text. On the common driver, $\Idep$ equals Hi-Fi while MMI adds $\delta=0.133$ bits to both atoms. On the collider, every PID with $\delta>0$ reports redundancy between two independent sources, from 0.068 bits ($\Idep$) to 0.208 bits (MMI). On the chain and the suppressor the segment has length zero and every nonnegative PID agrees with Hi-Fi, while $\Iccs$ produces a negative unique atom on the chain. On the collider with coupled sources, Hi-Fi reports net redundancy ($C=+0.035$ bits) while $\Idep$ finds 0.107 bits of synergy and MMI 0.361 bits: synergy that merely offsets redundancy is invisible to Hi-Fi.

\begin{table*}[t]
\caption{Gaussian triplets (population values, bits). $r_1=\mathrm{corr}(Y,X_1)$, $r_2=\mathrm{corr}(Y,X_2)$, $r_{12}=\mathrm{corr}(X_1,X_2)$, rounded to two decimals; the information quantities were computed from the exact correlations of the generating model. Measure columns give Red, Syn and, for $\Idep$ and $\Iccs$ where $\dmax>0$, the position $\lambda$ (MMI is at $\lambda=1$). MMI equals BROJA and $\Imin$; $\Iccs$ by Monte Carlo. The first six structures are specified directly by their correlations: common driver, $X_1\perp X_2\mid Y$ ($r_{12}=r_1r_2$); collider, $X_1\perp X_2$ ($r_{12}=0$); chain $Y\to X_1\to X_2$ ($r_2=r_1r_{12}$); suppressor, $r_2=0$; and the two coupled variants. Direct edge + confounder is the standardized linear model $X_1=0.8X_2+\varepsilon_1$, $Y=0.4X_1+0.8X_2+\varepsilon_2$, $\varepsilon_i\sim N(0,1)$ independent (Sec.~\ref{sec:mmi}).}
\label{tab:gauss}
\begin{ruledtabular}
\scriptsize
\setlength{\tabcolsep}{1.4pt}
\begin{tabular}{llccccccccc}
Structure & $r_1,r_2,r_{12}$ & $I_1,I_2,I_{12}$ & $C$ & $U_1/U_2$ & $R/S$ & $\dmax$ & MMI & $\Idep$ & $\Iccs$\\
\hline
\input{tables/tab_gauss}
\end{tabular}
\end{ruledtabular}
\end{table*}

Table~\ref{tab:ising} lists the Ising configurations of Sec.~\ref{sec:triadic}. The collider-like pairwise model has net synergy with $\Ithree=0$, and the pairwise-reference PID equals Hi-Fi there; in the mixed configuration, pairwise redundancy and triadic synergy nearly cancel in $C$.

\begin{table*}[t]
\caption{Ising triplets (bits). Hi-Fi reports the net co-information $C=\Itwo-\Ithree$; the other columns give (Red, Syn). BROJA and MMI coincide on all four configurations. The pairwise-reference PID (Sec.~\ref{sec:pairref}) reports $\pos{\Itwo}$ and $\Ithree+\pos{-\Itwo}$.}
\label{tab:ising}
\begin{ruledtabular}
\footnotesize
\begin{tabular}{lcccccccc}
Configuration & $\Itwo$ & $\Ithree$ & $C$ & Hi-Fi $R/S$ & $\dmax$ & $\Idep$ & BROJA & pairwise ref.\\
\hline
\input{tables/tab_ising}
\end{tabular}
\end{ruledtabular}
\end{table*}

Table~\ref{tab:mmi} gives the composition of MMI's synergy, Eq.~\eqref{eq:layers}, on the three families of ensemble C (Sec.~\ref{sec:mmi}).

\begin{table*}[t]
\caption{Composition of synergy in three simulated families (ensemble C). Columns 4--6 give the share of triplets in which MMI reports synergy, in which Hi-Fi reports net synergy ($S$), and in which MMI reports synergy while $S=0$ (synergy present: more than 1\% of $I_{12}$); binomial standard errors are at most 2.0 points for the binary families and 0.4 points for the Gaussian one. The mean synergy fraction is $\Syn/I_{12}$ averaged over triplets. The last column is the share of all MMI synergy that is overlap, $\sum\dmax/\sum\Syn_{\mathrm{MMI}}$, with bootstrap standard errors (2000 resamples) of 2.6, 2.3 and 0.3 points for the three rows; the triadic share of the pairwise-reference synergy quoted in Sec.~\ref{sec:mmi}, 71\%, has a bootstrap standard error of 2.3 points. PR: pairwise-reference PID. For Gaussian triplets BROJA equals MMI and the pairwise reference equals Hi-Fi (Proposition~\ref{prop:gauss}).}
\label{tab:mmi}
\begin{ruledtabular}
\footnotesize
\setlength{\tabcolsep}{4pt}
\begin{tabular}{p{2.7cm}ccccccc}
Family & $n$ & mean $\Ithree$ & MMI syn. & Hi-Fi $S>0$ & MMI syn., $S=0$ & mean $\Syn/I_{12}$ (\%): & overlap share\\
 & & (bits) & (\%) & (\%) & (\%) & MMI/BROJA/$\Idep$/PR/Hi-Fi & (\%)\\
\hline
\input{tables/tab_mmi}
\end{tabular}
\end{ruledtabular}
\end{table*}

%% file: tables/tab_gauss.tex
common driver & 0.80, 0.60, 0.48 & 0.737, 0.322, 0.870 & $+$0.189 & 0.548/0.133 & 0.189/0 & 0.133 & 0.322, 0.133 & 0.189, 0 (0.00) & 0.289, 0.100 (0.76) \\
collider & 0.60, 0.50, 0.00 & 0.322, 0.208, 0.679 & $-$0.150 & 0.322/0.208 & 0/0.150 & 0.208 & 0.208, 0.357 & 0.068, 0.218 (0.33) & 0.041, 0.191 (0.20) \\
chain & 0.80, 0.56, 0.70 & 0.737, 0.271, 0.737 & $+$0.271 & 0.466/0 & 0.271/0 & 0 & 0.271, 0 & 0.271, 0 & 0.335, 0.064 \\
suppressor & 0.50, 0.00, 0.60 & 0.208, 0, 0.357 & $-$0.150 & 0.208/0 & 0/0.150 & 0 & 0, 0.150 & 0, 0.150 & $-$0.002, 0.148 \\
c.\ driver + coupling & 0.60, 0.60, 0.80 & 0.322, 0.322, 0.368 & $+$0.275 & 0.047/0.047 & 0.275/0 & 0.047 & 0.322, 0.047 & 0.275, 0 (0.00) & 0.292, 0.017 (0.36) \\
collider + coupling & 0.65, 0.65, 0.30 & 0.396, 0.396, 0.757 & $+$0.035 & 0.361/0.361 & 0.035/0 & 0.361 & 0.396, 0.361 & 0.142, 0.107 (0.30) & 0.164, 0.129 (0.36) \\
direct edge + confounder & 0.65, 0.72, 0.62 & 0.398, 0.529, 0.636 & $+$0.291 & 0.107/0.238 & 0.291/0 & 0.107 & 0.398, 0.107 & 0.291, 0 (0.00) & 0.353, 0.062 (0.58) \\

%% file: tables/tab_ising.tex
pairwise, collider-like: $J_{1Y}\!=\!J_{2Y}\!=\!0.8$, $J_{12}\!=\!-0.55$ & $-$0.082 & 0 & $-$0.082 & 0/0.082 & 0.135 & 0.024, 0.106 & 0.135, 0.217 & 0, 0.082 \\
pairwise, redundant: $J_{1Y}\!=\!J_{2Y}\!=\!0.8$ & $+$0.145 & 0 & $+$0.145 & 0.145/0 & 0.202 & 0.145, 0 & 0.347, 0.202 & 0.145, 0 \\
pure triadic (noisy parity): $J_3\!=\!1.099$ & 0 & 0.531 & $-$0.531 & 0/0.531 & 0 & 0, 0.531 & 0, 0.531 & 0, 0.531 \\
pairwise redundant + triadic: $J_{1Y}\!=\!J_{2Y}\!=\!0.8$, $J_3\!=\!0.6$ & $+$0.087 & 0.069 & $+$0.018 & 0.018/0 & 0.288 & 0.117, 0.099 & 0.306, 0.288 & 0.087, 0.069 \\

%% file: tables/tab_mmi.tex
binary, additive logit & 599 & 0 & 86 & 38 & 46 & 17.5/15.2/7.2/6.9/6.9 & 68 \\
binary, logit with $X_1X_2$ term & 600 & 0.018 & 97 & 61 & 35 & 29.5/28.1/21.7/21.4/18.1 & 44 \\
Gaussian, uniform correlations & 19999 & 0 & 90 & 60 & 29 & 29.9/29.9/23.1/20.9/20.9 & 22 \\

%% file: app_three.tex
\section{Proofs for three or more sources}
\label{app:three}

\subsection{Proof of Theorem~\ref{thm:three}}

Let $f$ be nondecreasing with $f(\emptyset)=0$, write $f_A=f(A)$ without braces ($f_i$, $f_{ij}$, $f_{123}$), and let $\{i,j,k\}=\{1,2,3\}$ throughout. Put
\begin{equation*}
\begin{aligned}
a_{ij}&=\pos{C_{i,j}},& d_{ij}&=\pos{-C_{i,j}},\\
b_i&=\pos{C_{i,jk}},& p_i&=f_i-b_i,
\end{aligned}
\end{equation*}
so that $a_{ij}-d_{ij}=C_{i,j}$ and $p_i=\min\{f_i,f_{123}-f_{jk}\}\ge0$. Two identities will be used, both following from the definition of the co-information:
\begin{align*}
C_{i,j}+C_{k,ij}&=f_1+f_2+f_3-f_{123},\\
C_{i,j}+C_{i,k}-C_{i,jk}&=\textstyle\sum_lf_l-\sum_{l<m}f_{lm}+f_{123}.
\end{align*}
Both right-hand sides are symmetric in the three sources.

\paragraph*{Free values.}
Zero overlap on the six pairs (Fig.~\ref{fig:lattice3}(a,d)) means $\Red_{A,B}=\pos{C_{A,B}}$, that is, $I_\cap(\{i\}\{j\})=a_{ij}$ and $I_\cap(\{i\}\{jk\})=b_i$. Together with $I_\cap(\{A\})=f_A$ this fixes $I_\cap$ at 13 of the 18 nodes. Five values remain free: $r=I_\cap(\{1\}\{2\}\{3\})$, $t_i=I_\cap(\{ij\}\{ik\})$ and $u=I_\cap(\{12\}\{13\}\{23\})$. M\"obius inversion, going up the lattice, gives the atoms
\begin{align*}
\pi(\{1\}\{2\}\{3\})&=r,\\
\pi(\{i\}\{j\})&=a_{ij}-r,\\
\pi(\{i\}\{jk\})&=b_i-a_{ij}-a_{ik}+r,\\
\pi(\{i\})&=p_i,\\
\pi(\{12\}\{13\}\{23\})&=u-r-\textstyle\sum_lb_l+\sum_{l<m}a_{lm},\\
\pi(\{ij\}\{ik\})&=t_i-u-p_i,\\
\pi(\{ij\})&=f_{ij}-t_i-t_j+u,\\
\pi(\{123\})&=f_{123}-\textstyle\sum_{l<m}f_{lm}+\sum_lt_l-u.
\end{align*}
A nonnegative PID with zero overlap on all six pairs exists if and only if $r$, $t_1$, $t_2$, $t_3$ and $u$ can be chosen so that all 18 atoms are nonnegative. The atoms $\pi(\{i\})=p_i$ are nonnegative for every $f$. Of the others, the first three lines involve only $r$, the last three only $u$ and the $t_i$, and the atom of $\{12\}\{13\}\{23\}$ couples the two groups.

\paragraph*{Lower group.}
The atoms of $\{1\}\{2\}\{3\}$, $\{i\}\{j\}$ and $\{i\}\{jk\}$ are nonnegative if and only if
\begin{equation}
\max\Bigl\{0,\ \max_i\,(a_{ij}+a_{ik}-b_i)\Bigr\}\le r\le\min_{l<m}a_{lm}.
\label{eq:app_lower}
\end{equation}

\paragraph*{Upper group.}
Write $t_i=u+p_i+s_i$. The atoms of $\{ij\}\{ik\}$, $\{ij\}$ and $\{123\}$ are nonnegative if and only if $s_i\ge0$, $s_i+s_j\le c_{ij}$ with $c_{ij}=f_{ij}-p_i-p_j-u$, and $\sum_ls_l\ge\sum_{l<m}f_{lm}-f_{123}-\sum_lp_l-2u$. Nonnegative $s_i$ with $s_i+s_j\le c_{ij}$ exist if and only if all $c_{ij}\ge0$, and their sum can then reach $\min\{\tfrac12\sum c,\ \sum c-\max c\}$ but no more: adding the three constraints bounds the sum by $\tfrac12\sum c$, and adding two of them to $s_k\ge0$ bounds it by $\sum c-c_{ij}$ for each pair; the smaller bound is attained at $s_i=(c_{ij}+c_{ik}-c_{jk})/2$ when these three values are nonnegative, and otherwise at the point with $s_k=0$ for the $k$ opposite the largest $c_{ij}$ and $s_i=c_{ik}$, $s_j=c_{jk}$. Writing $R=\sum_{l<m}f_{lm}-f_{123}-\sum_lp_l-2u$ for the required sum, the two conditions $\tfrac12\sum c\ge R$ and $\sum c-c_{ij}\ge R$ read
\begin{equation*}
u\ge\sum_{l<m}f_{lm}-2f_{123}\qquad\text{and}\qquad p_k\le f_{123}-f_{ij},
\end{equation*}
the second of which holds automatically; hence the upper group can be made nonnegative if and only if
\begin{equation}
\sum_{l<m}f_{lm}-2f_{123}\le u\le\min_{l<m}\,(f_{lm}-p_l-p_m),
\label{eq:app_upper}
\end{equation}
and the same inequality shows that this interval is never empty.

\paragraph*{Coupling.}
The atom of $\{12\}\{13\}\{23\}$ requires $u\ge r+\sum_lb_l-\sum_{l<m}a_{lm}$. By Eq.~\eqref{eq:app_upper}, $u$ can be raised to $\min_{l<m}(f_{lm}-p_l-p_m)$, so the coupling can be met if and only if $r\le f_{ij}-p_i-p_j-\sum_lb_l+\sum_{l<m}a_{lm}$ for every pair $(i,j)$. Inserting $p$ and $C_{i,j}=a_{ij}-d_{ij}$, the right-hand side becomes
\begin{equation*}
\sigma_k=a_{ki}+a_{kj}+d_{ij}-b_k,
\end{equation*}
where $k$ is the source not in the pair.

\paragraph*{Conditions.}
A zero-overlap PID therefore exists if and only if some $r$ satisfies Eq.~\eqref{eq:app_lower} and $r\le\sigma_k$ for $k=1,2,3$, that is, if and only if
\begin{enumerate}
\item[(a)] $a_{ij}+a_{ik}-b_i\le a_{lm}$ for every source $i$ and pair $(l,m)$,
\item[(b)] $\sigma_k\ge0$ for every $k$,
\item[(c)] $a_{ij}+a_{ik}-b_i\le\sigma_k$ for every $i\ne k$.
\end{enumerate}
The remaining comparisons hold trivially, since $a_{lm}\ge0$ and $\sigma_i-(a_{ij}+a_{ik}-b_i)=d_{jk}\ge0$. Condition (b) states that the right-hand side of Eq.~\eqref{eq:dS} is nonpositive for $A=\{k\}$, $B=\{i\}$, $B'=\{j\}$, i.e.\ (S). Condition (a) for $(l,m)=(i,j)$ and $(i,k)$ reads $a_{ik}\le b_i$ and $a_{ij}\le b_i$, i.e.\ (M). It remains to show that (M) implies the other cases.

For (a) with $(l,m)=(j,k)$: if $a_{ij}=0$ or $a_{ik}=0$, the claim follows from (M). Otherwise $a_{ij}+a_{ik}-b_i\le C_{i,j}+C_{i,k}-C_{i,jk}$, which by the second identity equals $C_{j,i}+C_{j,k}-C_{j,ik}$. (M) for the ordered pair $(j,i)$ gives $b_j\ge a_{ij}=C_{i,j}>0$, hence $C_{j,ik}=b_j\ge C_{i,j}$, and the expression is at most $C_{j,k}\le a_{jk}$.

For (c), with $j$ the third source, the inequality reduces to $C_{i,j}+b_k\le a_{jk}+b_i$. If $b_k=0$, it follows from $C_{i,j}\le a_{ij}\le b_i$, which is (M). Otherwise $b_k=C_{k,ij}$, and by the first identity $C_{i,j}+C_{k,ij}=C_{j,k}+C_{i,jk}\le a_{jk}+b_i$.

Hence (M) and (S), for every ordering of the sources, are sufficient. The proof is constructive: $r=\max\{0,\max_i(a_{ij}+a_{ik}-b_i)\}$, $u=\min_{l<m}(f_{lm}-p_l-p_m)$ and $s$ at the maximum of its sum give a zero-overlap PID.

\subsection{Proof of Proposition~\ref{prop:identity}}\label{app:prop9}
For an identity target $I(Y;X_A)=H(X_A)$, so $C_{A,B}=H(X_A)+H(X_B)-H(X_{A\cup B})=I(X_A;X_B)\ge0$. Zero overlap on a pair therefore means $\Syn_{A,B}=0$, and with nonnegative atoms every synergy atom of the pair vanishes.

Every node $\alpha$ with an element $E$ of two or more sources is a synergy atom of a pair that Hi-Fi examines, namely $(\{e\},E\setminus\{e\})$ for any $e\in E$. The union $E$ lies in $\mathcal U(\alpha)$, but $\{e\}$ and $E\setminus\{e\}$ do not, because an element of $\alpha$ inside either of them would be a proper subset of $E$, which an antichain cannot contain. Zero overlap on the pairs that Hi-Fi examines thus leaves only the $2^n-1$ nodes $\alpha_T=\{\{t\}:t\in T\}$ whose elements are single sources.

The node $\alpha_T$ has up-set $\{A:A\cap T\neq\emptyset\}$, so Eq.~\eqref{eq:shannon_n} becomes $H(X_A)=\sum_{T\cap A\neq\emptyset}\pi(\alpha_T)$. Subtracting it from the equation for $A=[n]$ gives $H(X_B\mid X_{[n]\setminus B})=\sum_{\emptyset\neq T\subseteq B}\pi(\alpha_T)$ for every $B\subseteq[n]$, a zeta transform on the Boolean lattice. Its unique solution is the information diagram, $\pi(\alpha_T)=\mu(T)$~\cite{Yeung1991}. A nonnegative PID with zero overlap on the examined pairs therefore exists only if $\mu\ge0$, and it is unique.

Conversely, let $\mu\ge0$. The PID with $\pi(\alpha_T)=\mu(T)$ has no synergy atom on any coarse-grained pair: if a single source $\{t\}$ lies inside $A\cup B$, it lies inside $A$ or inside $B$. Its redundancy is $\Red_{A,B}=\sum_{T\cap A\neq\emptyset,\,T\cap B\neq\emptyset}\mu(T)=I(X_A;X_B)=\pos{C_{A,B}}$, so it has zero overlap on every coarse-grained pair, not only on those that Hi-Fi examines. For $n=3$, $\mu(\{i\})=H(X_i\mid X_j,X_k)$ and $\mu(\{i,j\})=I(X_i;X_j\mid X_k)$ are nonnegative, and only $\mu(\{1,2,3\})=C(X_1;X_2;X_3)$ can be negative.

\subsection{A four-source set function that satisfies (M) and (S) but admits no zero-overlap PID}
\label{app:n4}

Let $f(\{i\})=0$ for $i=1,\dots,4$, $f(\{2,4\})=0$, $f(A)=1$ for the other five pairs and for all triples, and $f([4])=2$. Then $f$ is nondecreasing, every co-information $C_{A,B}=f(A)+f(B)-f(A\cup B)$ is $0$ or $-1$, so every $\pos{C_{A,B}}$ vanishes and the right-hand sides of Eqs.~\eqref{eq:dM} and~\eqref{eq:dS} are nonpositive for every choice of disjoint $A,B,B'$: (M) and (S) hold. Suppose a nonnegative PID $\pi$ had zero overlap on all 25 coarse-grained pairs. Since every $C_{A,B}\le0$, zero overlap means $I_\cap(\{A\}\{B\})=0$ at every pair node, and by monotonicity of $I_\cap$ every atom whose antichain contains two disjoint elements vanishes. Moreover $f(\{i\})=0$ and $f(\{2,4\})=0$ force every atom whose antichain has an element inside $\{i\}$ or inside $\{2,4\}$ to vanish, by Eq.~\eqref{eq:shannon_n}. The surviving antichains consist of pairwise intersecting sets of size at least two, none of them $\{2,4\}$. For such an antichain $\alpha$ let $K(\alpha)=\bigcap\alpha$ and let $e(\alpha)$ be its number of two-element members. Equation~\eqref{eq:shannon_n} gives $\sum_\alpha\pi(\alpha)=f([4])=2$; for each $i$, $\sum_{\alpha:\,i\in K(\alpha)}\pi(\alpha)=f([4])-f([4]\setminus i)=1$, because the atoms counted in $I(Y;X_{[4]})$ but not in $I(Y;X_{[4]\setminus i})$ are those all of whose elements contain $i$; and for each of the five pairs $\{i,j\}\ne\{2,4\}$, $\sum_{\alpha:\,\{i,j\}\in\alpha}\pi(\alpha)=f(\{i,j\})=1$, because a surviving element inside $\{i,j\}$ can only be $\{i,j\}$ itself. Summing, $\sum_\alpha\pi(\alpha)\,[e(\alpha)+|K(\alpha)|]=5+4=9$, so the average of $e+|K|$ under the weights $\pi$ is $9/2$. But every surviving antichain has $e(\alpha)+|K(\alpha)|\le4$: a star $\{ij\}\{ik\}\{il\}$ has $3+1$, the triangle $\{ij\}\{ik\}\{jk\}$ has $3+0$, the top node $\{[4]\}$ has $0+4$, and every other intersecting antichain of sets of size at least two has a smaller value. This contradiction shows that no such $\pi$ exists; a linear program gives a minimal total overlap of $2$ over all 25 pairs, while zero overlap on the 22 pairs with a single source on one side is feasible.